\pdfoutput=1
\documentclass[fleqn,10pt]{wlscirep}

\title{Hi-C-constrained physical models of human chromosomes recover functionally-related properties of genome organization}

\author[1,*,+]{Marco Di Stefano}
\author[2,+]{Jonas Paulsen}
\author[3]{Tonje G. Lien}
\author[4,5,6]{Eivind Hovig}
\author[1]{Cristian Micheletti}

\affil[1]{SISSA, International School for Advanced Studies, Trieste, I-34136, Italy.}
\affil[2]{Institute of Basic Medical Sciences, University of Oslo, Oslo, 0317, Norway.}
\affil[3]{University of Oslo, Department of Mathematics, Oslo, 0316, Norway.}
\affil[4]{Institute for Cancer Research, Oslo University Hospital, Department of Tumor Biology, Oslo, 0310, Norway.}
\affil[5]{University of Oslo, Department of Informatics, Oslo, 0316, Norway.}
\affil[6]{Institute of Cancer Genetics and Informatics, Oslo, 0310, Norway.}

\affil[*]{marco.distefano@cnag.crg.eu}

\affil[+]{these authors contributed equally to this work}

\begin{abstract}
Combining genome-wide structural models with phenomenological data is at the forefront of efforts to understand the organizational principles regulating the human genome. Here, we use chromosome-chromosome contact data as knowledge-based constraints for large-scale three-dimensional models of the human diploid genome. The resulting models remain minimally entangled and acquire several functional features that are observed \emph{in vivo} and that were never used as input for the model. We find, for instance, that gene-rich, active regions are drawn towards the nuclear center, while gene poor and lamina-associated domains are pushed to the periphery. These and other  properties persist upon adding local contact constraints, suggesting their compatibility with non-local constraints for the genome organization. The results show that suitable combinations of data analysis and physical modelling can expose the unexpectedly rich functionally-related properties implicit in chromosome-chromosome contact data. Specific directions are suggested for further developments based on combining experimental data analysis and genomic structural modelling.
\end{abstract}
\begin{document}

\flushbottom
\maketitle
\thispagestyle{empty}

\section*{Introduction}
The advent of experimental techniques to study the structural organization of the genome has opened new avenues for clarifying the functional implications of genome spatial arrangement. For instance, the organization of chromosomes in territories with limited intermingling was first demonstrated by fluorescence in-situ hybridization (FISH) experiments~\cite{cremer2001,pombo2006} and, next, rationalised in terms of memory-effects produced by the out-of-equilibrium mitotic $\to$ interphase decondensation~\cite{grosberg1993,sikorav1994,plospaper,vettorel2009,bjpaper,lieberman2009,dorier2009}. These effects are, in turn, essential for the subsequent chromosomal recondensation step of the cell cycle~\cite{plospaper,bjpaper}. More recently, chromosome conformation capture techniques have allowed for quantifying the contact propensity of pairs of chromosome regions, hence providing key clues for the hierarchical organization of chromosomes into domains with varying degree of compactness and gene activity~\cite{Dekker:2002ka,lieberman2009,dixon2012,rao2014}.

Since their very first introduction\cite{Dekker:2002ka}, conformational capture experiments have been complemented by efforts to build coarse-grained models of chromosomes~\cite{langowski2010,mirny2011,nicodemi2014}. These modelling approaches have been used with a twofold purpose. On the one hand,  general models for long and densely-packed polymers have been used to compare their contact propensities and those inferred from Hi-C data. These approaches are useful to understand the extent to which the Hi-C-probed genome organization depends on general, aspecific physical constraints~\cite{grosberg1993,plospaper,mateos2009,lieberman2009,tanay2011,mirny2011,sextonHiCFly,barbieri2012,jost2014,nicodemi2014,giorgetti2014,serra2015}. On the other hand, Hi-C and other experimental measurements have been used as knowledge-based constraints to build specific, viable candidate three-dimensional representations of chromosomes\cite{Dekker:2002ka,duan2010,mirny2011,bau2011,kalhor2012}. These models are valuable because they can expose the genomic structure-function interplay to a direct inspection and analysis, a feat that cannot be usually accomplished with the sole experimental data\cite{Dekker:2002ka}.

Developing such models is difficult. In part, this is because it requires overcoming the limitations of the (currently unavoidable) dimensional reduction where a set of  contact propensities is measured in place of the actual three-dimensional conformations, and still obtain the latter. But an additional and key difficulty is the structural heterogeneity of the chromosomal conformational ensemble that is probed experimentally. As for the simpler, but still challenging, problem of proteins with structurally-diverse substates\cite{Dedmon:2005ck,Camilloni:2014iy}, such conformational heterogeneity makes it impossible for using all phenomenological restraints to pin down a unique representative structure, and suitable methods must be devised to deal with the inherent heterogeneity.

Here, by building on previous modelling efforts\cite{Dekker:2002ka,duan2010,mirny2011,bau2011,kalhor2012}, we tackle these open isssues and ask whether Hi-C data subject to a suitable statistical selection can be indeed be used as phenomenological constraints to obtain structural models of the complete human diploid genome that are  viable, i.e. that possess correct functionally-related properties.

The key elements of our approach are two. First, we use advanced statistical tools to single out local and non-local \emph{cis-}chromosome contacts that are significantly enriched with respect to the reference background of Hi-C data. Second, we employ steered molecular dynamics simulations to drive the formation of these constitutive contacts in a physical model of the human diploid genome, where chromosomes are coarse-grained at the 30nm level. The viability of this general strategy is here explored for two distinct human cell lines: lung fibroblasts (IMR90) and embryonic stem cells (hESC).

Various functional aspects of the genome organization have been previously addressed with structural models, see ref.~\cite{serra2015} for recent reviews. Some of these studies have addressed the architecture of specific, local functional domains~\cite{bau2011,scialdone2011,giorgetti2014}, including the formation and plasticity of topologically associated domains (TADs)~\cite{ledily2014,benedetti2014,sanborn2015,fudenberg2016,chiariello2016,tiana2016}. Structural models were also used to explore general features of the structure-function relationship, such as the interplay of gene co-expression and co-localization in human chromosome 19~\cite{distefano2013} or of epigenetic states and genome folding in {\it D. melanogaster}~\cite{jost2014}.

Other studies have instead dealt with the challenge of modelling entire yeast~\cite{ShRec3D,duan2010,wong2012,tjong2012,gong2014} or human chromosomes~\cite{kalhor2012,zhang2015} consistently with available experimental data, particularly for the spatial proximity of chromosome {\em loci}. The two main challenges of these approaches are: the use of suitable data analysis strategies for inferring pairwise distances from the phenomenological data, such as Hi-C~\cite{ShRec3D,duan2010,giorgetti2014,MeluzziArya,zhang2015,serra2016}, and the optimal use of the distances as phenomenological constraints for three-dimensional models \cite{duan2010,kalhor2012,giorgetti2014,zhang2015,serra2016}.

Our study addresses both issues and complements earlier efforts in several respects. For data analysis, we use a statistical test based on the zero-inflated negative binomial (ZiNB) distribution to identify significant entries (contacts) from Hi-C matrices of ref.~\cite{dixon2012} that are much enriched with respect to the expected (background) occurrence of contacts. Typically, this reference distribution is taken as the standard binomial one, which is parametrized based on genomic distance between loci, as well as various Hi-C technical biases\cite{heinz2010,duan2010,ay2014,mifsud2015}. In other selection schemes the conditional expectancy has been used to parametrize a negative binomial model based on the interaction frequencies between the restriction fragments\cite{jin2013}. Recently, Rao {\em et al}\cite{rao2014} have taken advantage of ultra high resolution Hi-C data to identify focal peaks in the Hi-C heatmaps by local scanning.

The advantageous property of the ZiNB scheme is its capability to deal with the inevitable sparsity of Hi-C matrices. The distinctive feature of our structural modelling is, instead, the seamless combination of the following features: the modelling is applied to the entire diploid genome inside the nucleus, the coarse-graining level is {\em uniformly} set to match the physical properties of the $30$nm fiber and, finally, steered molecular dynamics simulations are used to promote the formation of a subset of the Hi-C contacts, only the significant ones, allowing the unconstrained regions of the chromosomes to organize only under the effect of aspecific physical constraints.
The approach is also robust for the introduction of an independent set of constraints based on the high-resolution Hi-C measurements in ref.~\cite{rao2014}, which provide information about local interactions associated with the boundaries of topologically associating domains (TADs).

Using our approach, we found that the model chromosomes remain mostly free of topological entanglement and acquire various properties distinctive of the {\em in vivo} genome organization. In particular, we found gene-rich and gene-poor regions, lamina-associated domains (LADs), \emph{loci} enriched in histone modifications, and Giemsa bands to be preferentially localized in the expected nuclear space.
To our knowledge, this study, which builds on and complements previous genome modelling efforts \cite{duan2010,ShRec3D,serra2015} is the first to engage in genome-wide physical modelling for two different human cell lines, based on Hi-C data from two different groups, and processed with two alternative statistical analyses. While this breadth ought to make the results interesting {\em per se}, the fact that several correct functional features are systematically recovered, makes the approach more relevant and useful for genome modelling. In fact, besides providing a concrete illustration of the genomic structure-function interplay, the results prompt the further development of coarse-grained models as an essential complement of experimental data analysis. Specific directions for such advancements are suggested.

\section*{Results}
\subsection*{Significant pairwise constraints from Hi-C data.} As input data for the knowledge-based three-dimensional (3D) modelling of human chromosomes, we used Hi-C measurements from  lung fibroblasts (IMR90) and embryonic stem cells (hESC)~\cite{dixon2012}. These data sets provide genome-wide 3D contact information between chromosome regions at $100$ kilobases (kb) resolution. We focused on {\em cis}-chromosome Hi-C contacts, which, in contrast to {\em trans}-chromosome ones, show rich and robust pairing patterns~\cite{lieberman2009}. The matrix of {\em cis} contacts is sparse as most of the {\em a priori} possible pairings have no associated reads, either because they are genuinely not in spatial proximity, or because their contacting probability is too low to be reliably detected for a given sequencing depth.

This data sparsity must be appropriately dealt with for pinpointing the statistically-significant {\em cis}-chromosome pairings that serve as knowledge-based constraints. To this end, we carried out a stringent statistical analysis using the \emph{zero-inflated negative binomial}  distribution (see Methods).

We accordingly singled out $16,409$ and $14,928$ significant pairings for IMR90 and hESC cells, respectively, using a $1\%$ threshold for the false-discovery rate, see Supplementary Tables S$1$ and S$2$, and Supplementary Fig. S$1$.
The number of significant contacts is comparable, and actually larger by a factor of 2, than those found by Rao {\em et al.} using different selection criteria and different Hi-C data (and that we shall later incorporate in our modelling as well).
The significant pairings obtained with this culling procedure ought to correspond to a core of contacts that are likely present across the heterogenenous conformational repertoire populated by chromosomes. Therefore, this core does not include all contacts present in actual chromosome conformations, so that the restrained model structures are expectedly underconstrained with respect to the real system. Still, as we show later, these core contacts suffice to correctly pin down the larger scale functional features of chromosome organization.

For both cell lines the number of significant pairings correlates only weakly with chromosome length ($p$-value$>0.08$ of non-parametric Kendall rank correlation), but correlates significantly with the number of genes in the chromosomes ($p$-value$<0.005$), see Supplementary Fig. S$2$.  Consistent with this observation, the highest linear density of significant contacts is found for the chromosomes $19$ and X, where the gene density is high, while the lowest is found for chromosomes $13$ and $18$, where the gene density is low.

The observed correlation is not obvious {\em a priori}. In fact, because the cross-linking step in Hi-C experiments is not specific for gene-rich regions, the resulting contacts are expected to be unbiased in this respect. In addition, we used the normalization method of Imakaev \emph{et al.}~\cite{imakaev2012} to correct for various technical biases, including the possible difficulty of mapping reads on gene-poor chromosomes, whose sequence repetitiveness can be high. It is therefore plausible that the statistical selection criterion is capable of singling out those contact patterns that, being significantly enhanced across the probed cell population, are relevant for gene function.

\subsection*{Genome-wide models from spatial constraints}
The statistically significant Hi-C pairings were used as target contacts for the model diploid system of human chromosomes. Following refs.~\cite{plospaper,bjpaper,distefano2013}, each chromosome was modelled as a semi-flexible chain of beads~\cite{kremer_jcp} with $30$ nm diameter, corresponding to about $3$ kb~\cite{30nm}. Two copies of each autosome plus one of the X chromosome were packed at the nominal genome density inside a confining nuclear environment. For simplicity, the nuclear shape has been chosen to be spherical with a radius of $4,800$ nm, neglecting the flattened ellipsoidal shape of fibroblast cells~\cite{Bolzer2005}. The initial positions of the chromosomes were assigned in a stochastic way based on the phenomenological radial position propensities of ref.~\cite{Bolzer2005} (phenomenologically placed chromosomes). Steered molecular dynamics simulations were next used to promote the formation of contacts corresponding to the significant Hi-C pairings. Notice that the heterogeneity of the Hi-C sample should make it unfeasible to satisfy simultaneously all contacts corresponding to significant Hi-C pairings. Rather, the selected pairings ought to consist of incompatible subsets of feasible contacts.

The steering process was repeated independently $10$ times for each considered cell line. To ensure the statistical independence, we considered one conformation per run, namely the snapshot taken at the end of the steering protocol, for all the following structural analysis. For simplicity and definiteness, we mostly focus on lung fibroblasts (IMR90) with phenomenologically placed chromosomes and point out the relevant analogies with the human embryonic stem cells (hESC) in specific contexts. These and other reference cases, such as pre-steering models and steered systems with random initial placements of the chromosomes, are detailed in Supplementary Fig. S$3$ and S$5$-S$16$.

\begin{figure}[ht]
  \centering
  \includegraphics[width=\textwidth]{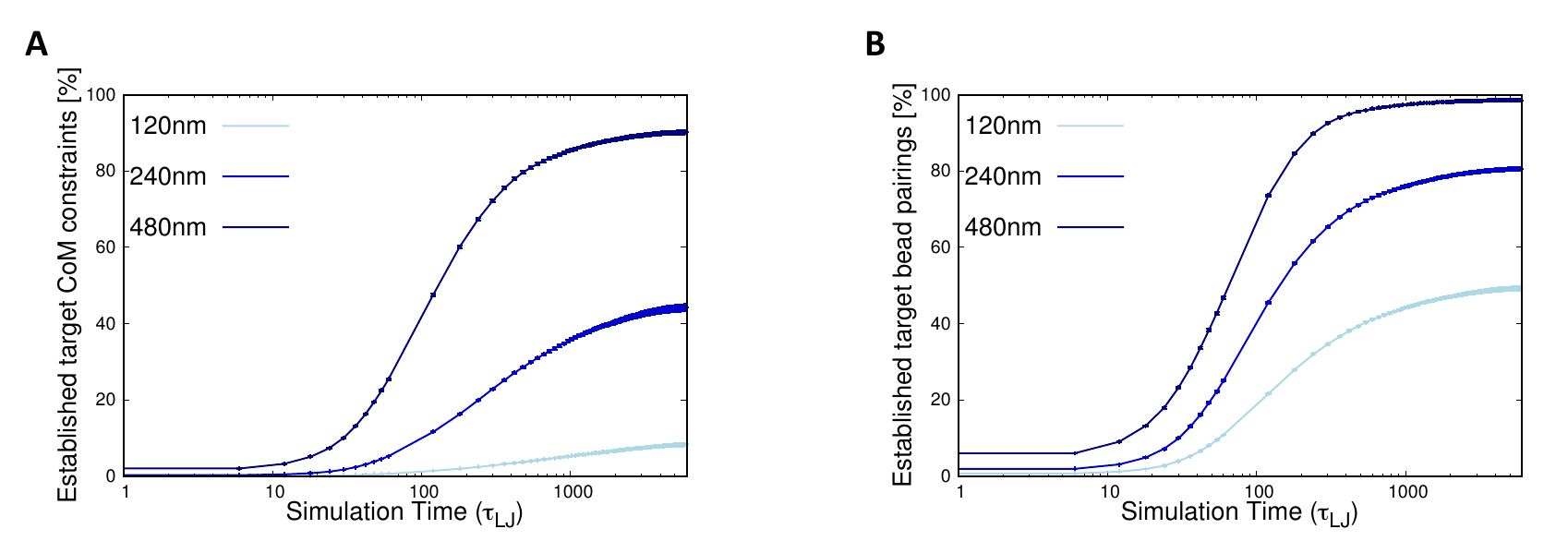}
  \caption{{\bf Evolution of the satisfied target constraints.} The
    curves show the increase of the percentage of target contacts that
    are established in the steering dynamics of the model IMR90
    genome. Different proximity criteria for defining contacts are
    used for the two panels: ({\bf A}) proximity of the centers of
    mass of the target regions, which are $100$ kb-long and span
    $33$ beads, ({\bf B}) proximity of the closest pair of beads in
    the target regions. The latter contact definition is arguably
    closer in spirit to the chromosome pairings which, quenched by
    ligation, contribute to Hi-C contacts. For each panel, the curves
    correspond to various cutoff distances: $120$ nm, $240$ nm and
    $480$ nm.}
  \label{fig:constraints}
\end{figure}

The data in Figures~\ref{fig:constraints}A and B show that the target proximity constraints, despite being practically all unsatisfied before steering, are progressively established in significant proportions during the steering process. The result is consistent with previous findings~\cite{distefano2013} and proves {\em a posteriori} that a large fraction of the selected constraints can indeed be simultaneously established in a three-dimensional model without being hindered by the physical incompatibilities within the target contacts. The same properties hold also in the cases of hESC cells and of random initial placements of the chromosomes (Supplementary Fig. S$3$).

\begin{figure}[ht]
  \centering
  \includegraphics[width=\textwidth]{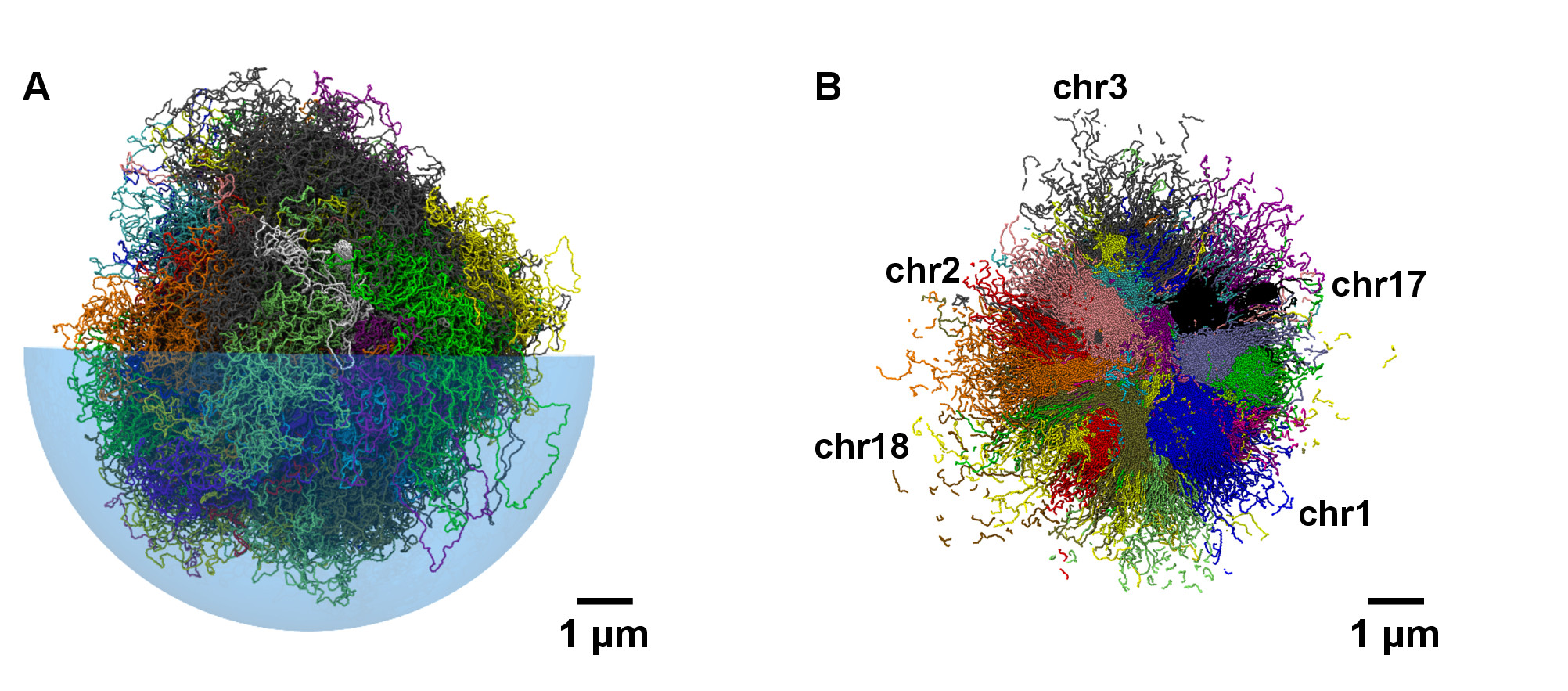}
  \caption{{\bf Spatial arrangement of chromosomes based on Hi-C data
      in lung fibroblast cells (IMR90).} ({\bf A}) Typical chromosomal
    spatial arrangement obtained applying the initial phenomenological
    radial positioning of ref.~\cite{Bolzer2005} and the steering
    dynamics based on the Hi-C data in
    ref.~\cite{dixon2012}. Different colors are used for different
    chromosomes. For visual clarity only one half of the enveloping
    nuclear boundary is shown, and it is rendered as a transparent
    hemisphere. ({\bf B}) Tomographic cut of the chromosomal system
    shown in panel A. The planar cut has a thickness of $150$nm.}
  \label{fig:snapshots}
\end{figure}

A typical arrangement of the steered chromosome conformation is shown in Figure~\ref{fig:snapshots}A. The accompanying tomographic cut (Figure~\ref{fig:snapshots}B) shows that chromosomes have a convex shape, as typically observed in FISH imaging~\cite{cremer2001,Bolzer2005}. As discussed later, the limited {\em trans-} intermingling observed in chromosomes at the initial, decondensed states, is preserved during steering and is present in the optimized chromosome conformations.

We compared the post-steering distance matrices with the matrices of Hi-C reads. This comparison is meant as a further {\em a posteriori} assessment of the system compliance to follow the actual constraints and acquire a spatial organization compatible with the full Hi-C matrices. To do so we used the non-parametric Kendall association test between corresponding entries of the two matrices. The results are shown, for all chromosomes, as Supplementary Fig. S$4$, and indicate a systematic significant correlation. This is actually an anticorrelation because shorter spatial distances reflect in higher number of reads. To better capture the interplay of aspecific (short-range) and specific (longer-range) domain organization the Kendall correlation coefficient was computed over corresponding entries with genomic distance larger than a minimum threshold, that was varied from 1Mbp to half the chromosome length. The resulting correlation profiles in Supplementary Fig. S$4$ show that post-steering distance matrices typically maintain a significant anticorrelation with Hi-C upon increasing the genomic distance threshold. By contrast, the same correlation, but measured for the initial chromosome distance matrix degrades rapidly with increasing threshold.
This is correct, because it reflects the increasing weight of longer-range specific interactions at the expense of the aspecific, shorter-range ones, of the initial state.
Chromosome specific features are hence systematically reproduced only by the steered model chromosomes.

\begin{figure}[ht]
  \centering
  \includegraphics[width=0.5\textwidth]{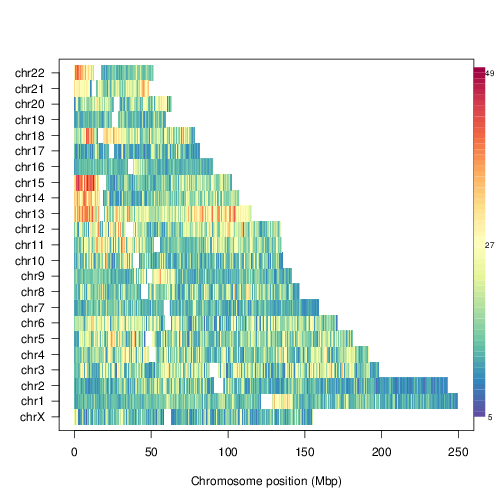}
  \caption{{\bf Genome-wide variability of radial bead position in
      lung fibroblast cells (IMR90) nuclei.} Numbers indicate the
    standard deviation of the radial position across the $10$
    replicate simulations. Data refer to phenomenologically
    prepositioned initial chromosomal locations. For a comparison
    with: the pre-steering case, alternative initial positionings and
    hESC cell lines, see Supplementary Fig. S$5$-S$8$.}
  \label{fig:IMR90_position_variability}
\end{figure}

\noindent{\bf Local variability of chromosomal nuclear positioning.}
The nuclear position variability of the IMR90 optimised chromosomes is shown in Figure~\ref{fig:IMR90_position_variability}, where the heatmap represents the standard deviations of the radial positioning of all chromosome portions.
Chromosomes $19$ and $X$ which have the highest number of genes, have the lowest average position variability, while acrocentric chromosomes $13$, $14$, $15$, $21$ and telomeric regions of chromosomes $18$, $21$ and $22$ have the largest one. At the same time, the observed position variations are plausible from the functional point of view, particularly regarding the increased motility found to occur for telomeric regions, compared to other, internal repeat regions in the human genome~\cite{wang2008,Robin2014}.

\subsection*{Nuclear positioning of functional regions.} The steering optimization of the genome-wide model is based on two phenomenological inputs, the significant Hi-C contacts and the typical radial placement of chromosomes, which are not simply nor manifestly connected to functional aspects. Therefore, a relevant question is whether functionally-related properties can at all be recovered and exposed by the optimized chromosome conformations.

We accordingly considered the radial placement, before and after steering, of gene-rich and gene-poor regions, of lamina-associated domains (LADs), of \emph{loci} enriched in H3K4me3 (activating), H3K9me3 and H3K27me3 (repressive) histone modifications. We also investigated the preferential nuclear localization of Giemsa (G)-bands, which are the resulting coloring patterns of the low-resolution chromosomal staining technique Giemsa banding. Using this technique, heterochromatic, AT-rich and relatively gene-poor regions are depicted by darkly (positive) staining bands, while less condensed chromatin, tending to be GC-rich and more transcriptionally active, appears as light (negative) bands.

\begin{figure}[ht]
  \centering
  \includegraphics[width=0.5\textwidth]{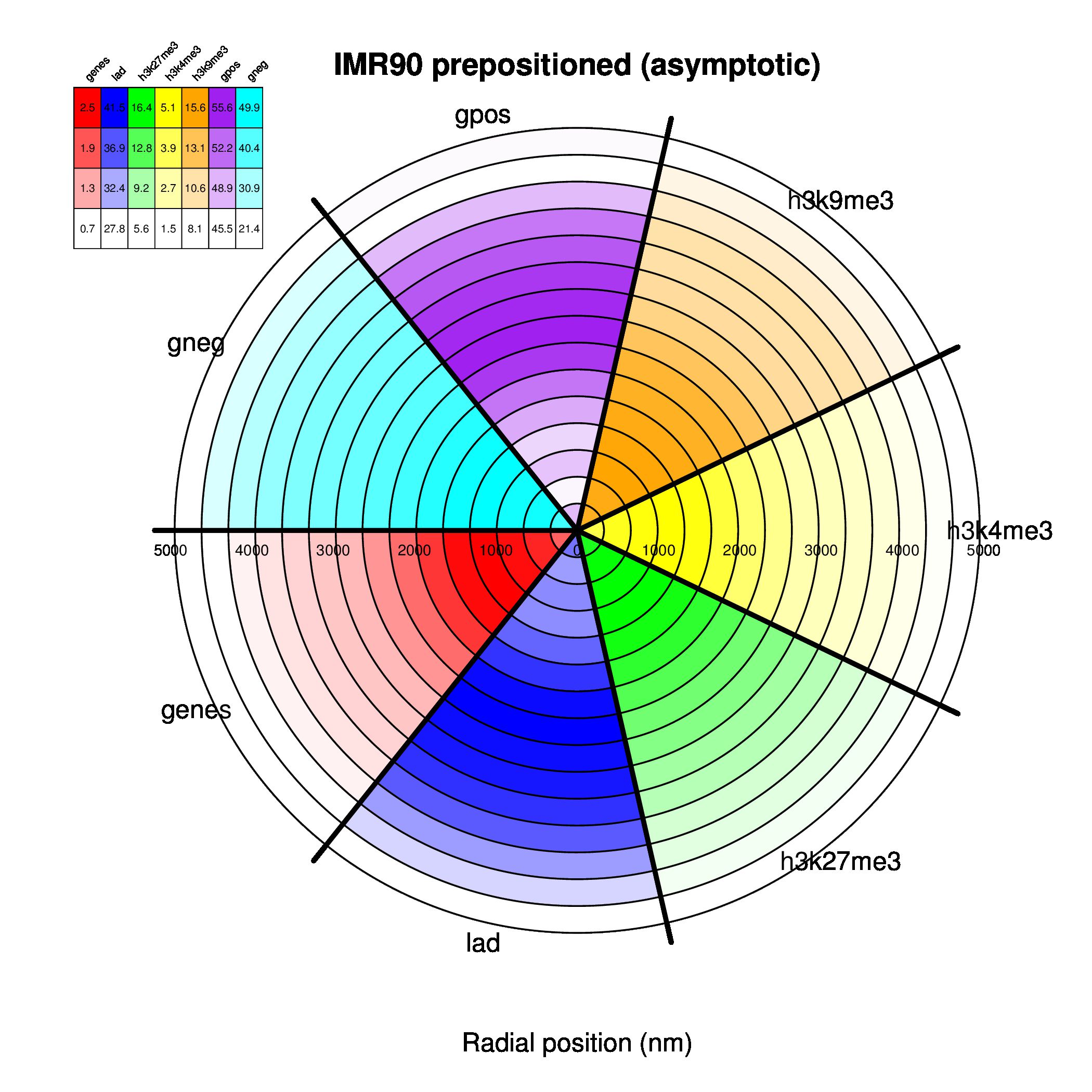}
  \caption{{\bf Nuclear positioning of functionally-related genomic
      regions in lung fibroblast cells (IMR90).} The central
    histogram gives the relative density percentage based on H3K9me3
    (orange), H3K4me3 (yellow), H3K27me3 (green), LADs (blue) and genes
    (red), and negative (cyan) and positive (purple) Giemsa staining
    bands. Circular slices indicate radial position (in nm)
    within the bounding nucleus aggregated across all $10$ replicate
    simulations. The legend indicates the percentage of beads associated
    with the given feature relatively to the total number of beads in
    the given radial shell.}
  \label{fig:IMR90_histograms}
\end{figure}

These all acquired the correct preferential positioning after steering. Specifically, chromosome regions that are rich in genes, associated with activating or repressing histone modifications or positive Giemsa staining bands occupy preferentially the nuclear center. Conversely, regions poor in genes or corresponding to positive Giemsa bands occupy preferentially the periphery. This holds for LADs too, whose simultaneous location at the nuclear periphery is not expected, given the highly dynamic association with the lamina {\em in vivo}~\cite{kind2013}. These properties  are shown in Supplementary Fig. S$13$-S$16$ for hESC, while for fibroblasts, where territorial radial positioning is known to be less definite~\cite{federico2008}, are shown in Figure~\ref{fig:IMR90_histograms}.

We found that: (i) prior to steering, genes are not preferentially near the nucleus center, and that (ii) for entirely random initial chromosome placements, the steered LADs locations are less peripheric. The significance of these differences are shown in Supplementary Fig. S$9$-S$12$ and indicates that the observed functional properties emerge specifically after introducing the phenomenological constraints on the genome structural model.

\begin{figure}[ht]
  \centering
  \includegraphics[width=\textwidth]{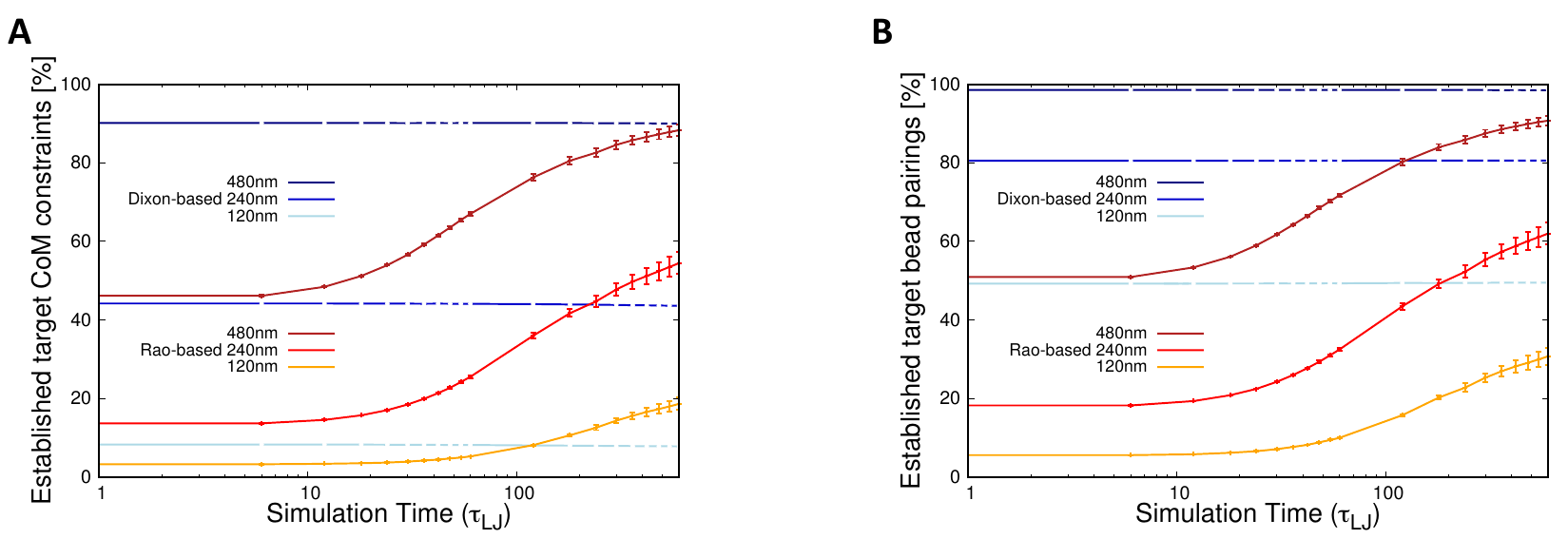}
  \caption{{\bf Evolution of the satisfied target constraints during the additional steering dynamics.} The curves show the percentage of the mantained target contacts from ref.~\cite{dixon2012} and the established new ones from ref.~\cite{rao2014} during the additional steering dynamics. The same contact criteria of Figure~\ref{fig:constraints} are used for the two panels: ({\bf A}) proximity of the centers of mass of the target regions, which are $100$ kb-long and span $33$ beads, ({\bf B}) proximity of the closest pair of beads in the target regions. For each panel, the curves correspond to various cutoff distances: $120$ nm, $240$ nm and $480$ nm.}
  \label{fig:constraints_Rao}
\end{figure}

The robustness of the functionally-related properties is implied by their consistency across the IMR90 and hESC cell lines. But it is best illustrated by the persistence of the same features upon adding an independent set of spatial constraints. In fact, after completing the steering with the significant IMR90 contacts from ref.~\cite{dixon2012}, we added those selected in ref.~\cite{rao2014} for  the same cell line, but in different, and higher resolution {\em in situ} Hi-C experiments. These additional target contacts are fewer in number ($8,040$) than the first set and are more local. In fact, the median sequence separation of contacting pairs is $220$ kb in the selected set of ref.~\cite{rao2014} while it is equal to $46.8$ Megabases (Mb) for our reference set. Within our top-down approach, where the constraints are used to introduce progressively detailed structural features on top of an initially  generic chromosome model, it is natural to apply the more local constraints of ref.~\cite{rao2014} after those from ref.~\cite{dixon2012}.
As it is shown in Figure~\ref{fig:constraints_Rao}, the added set has the same compliance to steering as the reference one, and the formation of the new contacts does not compete with or disrupt the former.
In fact, all previously-discussed  functionally related features persist with the added constraints, see Supplementary Fig. S$17$-S$19$. Interestingly, this compatibility suggests that organizational mechanisms at both local- and non-local levels, meaning sequence separations smaller than 220 kb or larger than 46.8 Mb, can simultaneously concur to the formation of large-scale genome topology and is consistent with current views of how the genome acquires the organization in local domains (TADs)~\cite{ciabrelli2015}.

\begin{figure}[ht]
  \centering
  \includegraphics[width=\textwidth]{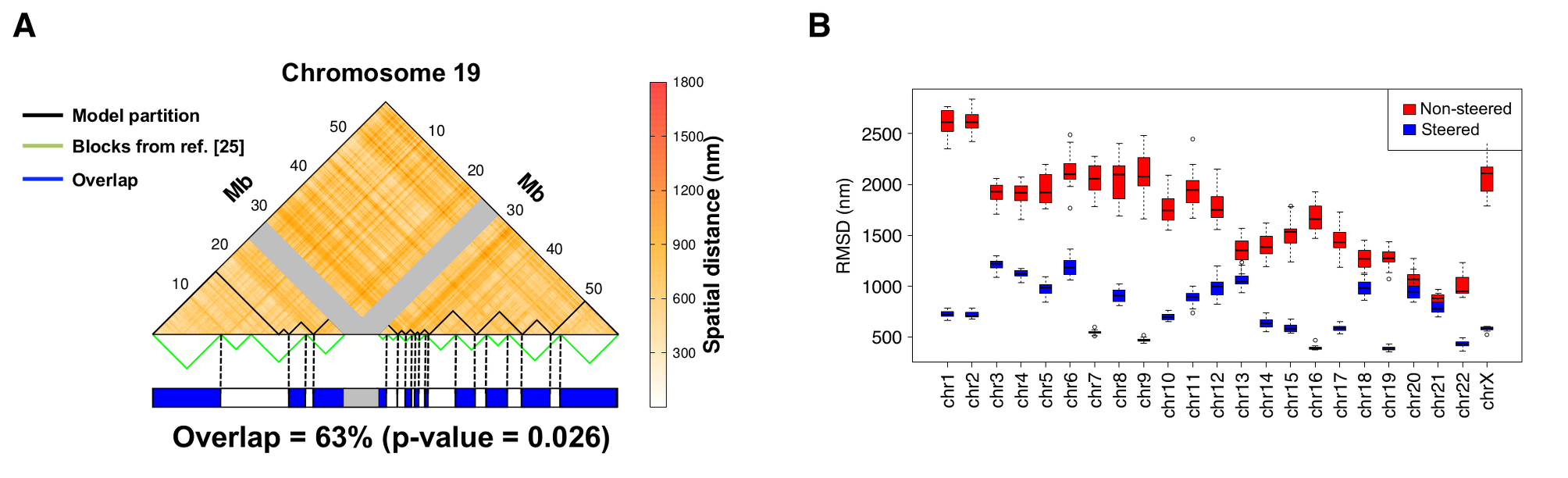}
  \caption{{\bf Comparison of large-scale chromosome features.} ({\bf A}) The upper triangle of the map of average spatial distance between $100$kb regions on chromosome $19$ at the end of the steering dynamics is shown. The gray bands mark the centromeric region. The boundaries of the $13$ spatial macrodomains, identified with a clustering analysis of the distance matrix (see Methods), are overlaid on the map and the boundaries of the spatial \emph{blocks} from ref.~\cite{kalhor2012} are shown below. The consistency of the two partitions is visually conveyed in the chromosome cartoon at the bottom. Overlapping regions, shown in blue, account for $63\%$ of the chromosome (centromere excluded). ({\bf B}) The RMSD of our chromosome models and the models inferred using the method in ref.~\cite{ShRec3D} are shown for non-steered (red) and steered conformations using target contacts based on ref.~\cite{dixon2012} (blue). The similarity of the two models is very clearly increased by the constrained steering procedure, and particularly so for chromosomes with a denser set of constraints (chr19, chr17, chr16, and chr7).}
  \label{fig:macrodomain_partition}
\end{figure}

In this regard, we have compared the macrodomain organization of the optimised chromosome $19$ with that obtained by Kalhor {\em et al.}~\cite{kalhor2012} based on an independent set of experimental proximity measurements. We chose chromosome $19$, because it has the highest linear density of imposed target constraints ($\sim 15$ constraints/Mb, see Supplementary Table S$3$), and all of them are simultaneously established in the optimally constrained models within $480$nm. For this comparison, we first used a clustering procedure (see Methods)  to optimally partition the chromosome arms into the same number of domains established in ref.~\cite{kalhor2012}. Next, we compared the overlap of the two domain partitions and established its significance by comparing it with the overlap distribution of random subdivisions of the two arms in the same number of clusters. The comparison, visualised in Figure~\ref{fig:macrodomain_partition}A, shows that the domains of chromosome 19 optimised by the combined constraints of ref.~\cite{dixon2012,rao2014} have a significant overlap $q=63\%$ ($p$-value = $0.026$) with those of ref.~\cite{kalhor2012}. Interestingly, the overlap is appreciably smaller before the addition of the constraints from ref.~\cite{rao2014}, $q=55\%$ ($p-value=0.065$), see Supplementary Fig. S$20$.\\
The results give a very vivid example of how the addition of independent sets of phenomenological constraints is not only viable (meaning that they do not interfere negatively) but actually allows to better expose the genuine large-scale organizational feature of the genome. The total number of combined constraints used for chr19 is $1,100$, equivalent to $\sim 19$ constraints per Mbp. This density of constraints with mixed local and non-local character thus appears to be a good target for robust chromosome modelling.

Furthermore, we compared the structural models, optimised with the phenomenological constraints based on Dixon {\em et al.} Hi-C data, with  the multidimensional scaling (MDS) constraints provided  by the analysis of Lesne \emph{et al.}~\cite{ShRec3D} of the very same Hi-C matrix. MDS is an approximate strategy for ``inverting'' of a well-populated matrix of pairwise-distances, and hence its applicability to genomic contexts required  downsampling the input Hi-C matrix. We accordingly lowered the structural resolution of our models to the $100$kb level (by taking the average bead position of all beads in each $100$kb bin), and then used the similarly resampled Dixon {\em et al.} Hi-C matrix to obtain the MDS model of each chromosome via the Floyd-Warshall algorithm~\cite{ShRec3D}. We compared both steered and non-steered chromosomes with the corresponding MDS-optimized structures, using the root mean square deviation (RMSD) after Procrustes superimposition. As Figure~\ref{fig:macrodomain_partition}B shows, the steered structures show a much greater similarity to the MDS-optimized structures, indicating that global organizational patterns of individual chromosomes are similar. Furthermore, chromosomes with a denser set of constraints (e.g. chr19, chr17, chr16, chr7) are more similar in terms of RMSD, again indicating that the comparison is sound. We emphasize that MDS is a method for coarse-grained optimization of smaller structures (e.g. single chromosomes), while our modelling approach allow for the joint modelling of all chromosomes simultaneously, and can account for their structural heterogeneity.

\subsection*{Chromosome pre-mitotic recondensation.} During the interphase $\to$ mitotis step of the cell cycle, the decondensed interphase chromosomes reconfigure in its characteristic rod-like shapes. This rearrangement, while certainly being assisted by topoisomerases, is also aided by the limited incidence of \emph{cis} and \emph{trans} topological constraints. This feature is, in turn, fostered by the out-of-equilibrium characteristics of the cell cycle~\cite{grosberg1993,sikorav1994,plospaper,bjpaper,vettorel2009,lieberman2009,mirny2011}. In fact, the observed chromosomal entanglement is significantly lower than for equivalent mixtures of long equilibrated polymers, which cannot reconfigure over biologically relevant time scales~\cite{plospaper}.

We tested the reconfiguration compliance of the optimised chromosome configurations by switching off the phenomenological target constraints after steering and replacing them with alternative target pairings
between {\em loci} at the regular sequence separation of $200$ kb. These constraints were chosen \emph{ad hoc} to promote the rearrangement into a linear succession of loops, analogously to the string-like (mitotic) chromosome models of refs.~\cite{sikorav1994,naumova2013}.

\begin{figure}[ht]
  \centering
  \includegraphics[width=\textwidth]{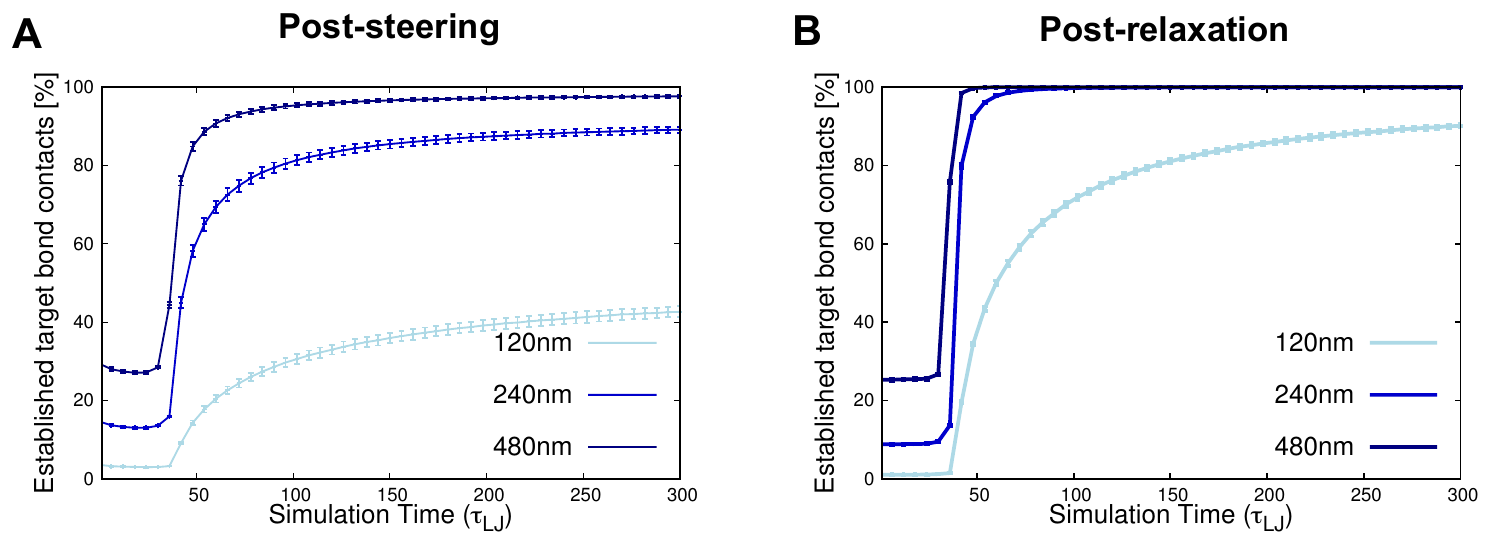}
  \caption{{\bf Time-evolution of the satisfied recondensation constraints.} The panels show the increase of the percentage of target contacts that are established during the recondensation dynamics starting from the optimally-steered conformations ({\bf A}) and the relaxed conformations ({\bf B}). As in Figure~\ref{fig:constraints} the curves correspond to various cutoff distances: $120$ nm, $240$ nm and $480$ nm. We notice that the compliance to the steering is similar in the two different starting conditions.}
  \label{fig:recondensation}
\end{figure}

\begin{figure}[ht]
  \centering
  \includegraphics[width=\textwidth]{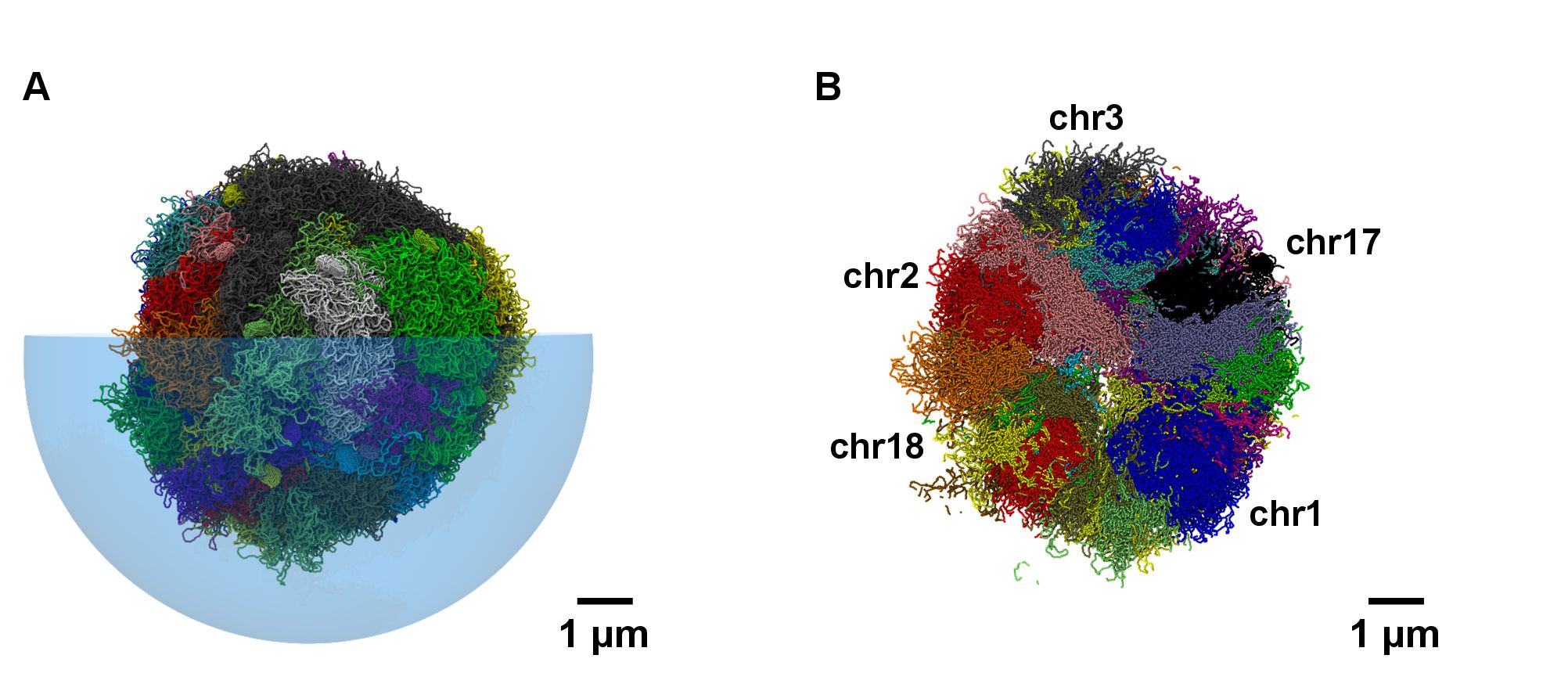}
  \caption{{\bf Chromosome segregation upon recondensation.} The
    chromosome conformations for lung fibroblast cells (IMR90)
    obtained using the initial phenomenological radial placement of
    ref.~\cite{Bolzer2005}, and the steering dynamics based on the
    Hi-C data in refs.~\cite{dixon2012} and ~\cite{rao2014} (see
    Figure~\protect{\ref{fig:snapshots}}A) were recondensed towards a
    mitotic-like arrangements by means of attractive interactions
    between pairs of {\em loci} (single beads) equally-spaced along
    the sequence at $200$ kb ($66$ beads).}
  \label{fig:snapshot_recondensation}
\end{figure}

The steered chromosomes were indeed able to reconfigure and establish most of the new constraints.
Their compliance to the target rearrangement is, in fact, very similar to the compliance of the pre-steering conformations which, having been relaxed from the initial rod-like arrangement, are ideally primed to be reconfigured efficiently, see  Figure~\ref{fig:recondensation}. The lack of significant topological barriers allows the condensing chromosomes to segregate neatly, see Figure~\ref{fig:snapshot_recondensation}, in qualitative accord with FISH observations~\cite{cremer2001,Bolzer2005}.

\section*{Discussion}
We studied the genome organization of human lung fibroblast (IMR90) and embryonic stem cells (hESC) by combining advanced statistical analysis of Hi-C data and coarse-grained physical models of chromosomes. The models were specialised for each cell line with phenomenological pairwise contacts corresponding to statistically-significant entries of the \emph{cis}-chromosome Hi-C heatmaps of ref.~\cite{dixon2012}. Being shared by an appreciable fraction of the cells probed experimentally, these contacts, despite not covering all contacts in actual chromosome conformations, ought to be simultaneously compatible, and physically viable. Their formation in the otherwise general models was promoted with steered molecular dynamics simulations on $10$ independent replicates per cell line.

With the combined data-analysis and modelling strategy, we studied whether relevant aspects of the genomic structure-function relationship could be retrieved from Hi-C data. For this genome-wide study, we use general chromosomes models that are discretised at the 30nm level.
This fine intrinsic granularity sets a lower bound for the optimised chromosome structure resolution. The latter, in fact, depends on the abundance and type of phenomenological pairwise contacts that are used as phenomenological constraints. The present approach, therefore, aptly complements previous efforts based on larger-scale models incorporating observations from different phenomenological sources (tethered conformation capture techniques~\cite{kalhor2012}).

We found that, by solely promoting the formation of statistically-significant Hi-C contacts, the model chromosomes, despite being underestrained compared to actual chromosomes, still acquire a number of properties that are distinctive of the {\em in vivo} organization. These properties cannot be simply inferred from the direct analysis of Hi-C maps only, a fact that stresses and reinforces the early intuition that three-dimensional models can best expose properties that are otherwise encoded only indirectly in pairwise contact matrices\cite{Dekker:2002ka}.

Specifically, in the optimised chromosome models gene-rich regions and activating or repressing histone modifications preferentially occupy the central space of the nucleus. By contrast, gene-poor ones and LADs occupy the nuclear periphery. All these features are in accord with experimental observations and with the functionally-related implications of the radial positioning of these regions. The localization of gene-rich regions in the nuclear center can, in fact, be an important prerequisite to organize them in clusters of proximal {\em loci} (transcriptional {\em foci}) providing an efficient means for their co-expression and co-regulation~\cite{takizawa2008}. Further, a likely explanation for these observations is that the gene-rich, centrally localized chromatin is most actively regulated via histone modifications, while the peripheral regions of the genome, which are less accessible, are preferentially constituted by repressed chromatin. In support of this view, it was recently proposed that lamina-associated chromatin compartments are depleted of most other histone modification signatures~\cite{rao2014}.

Furthermore, chromosomal regions associated with positive or negative Giemsa bands tend to occupy the nuclear periphery or centre, respectively. This result is again in accord with previous findings~\cite{federico2008}. It may also be correlated with the preferential placement of gene-rich and gene-poor regions as suggested in ref.~\cite{bickmore2013}. In fact, GC content in mammals, which is also reflected in the Giemsa banding patterns, is correlated with several genomic features that are potentially relevant from a functional viewpoint, including gene density, transposable element distribution, methylation levels, recombination rate, and expression levels. Thus, correlation studies will often tend to observe these together~\cite{PMID20530252}. It is however, also possible that Hi-C experimental uncertainties contribute in part to the observed effect~\cite{gavrilov2014,cheung2011}.

The robustness of the structure-function relationship recovered by using the IMR90 constraints from the data of ref.~\cite{dixon2012}, was tested by adding a further set of phenomenological constraints. These corresponded to the significant contacts selected in ref.~\cite{rao2014} from high-resolution {\em in situ} Hi-C experiments on the same cell line. The concomitant steering  with the two sets of constraints did not ruin the previously-established contacts and, in fact, acquired a sizeable proportion of the added ones. These facts indicate that the two sets are compatible, arguably because the significant contacts of ref.~\cite{rao2014}  are more local than those selected from the experimental data of ref.~\cite{dixon2012} with the different statistical methods described in the Methods section. As a result, the two sets complement each other for pinning the large- and small-scale structural features of the doubly-steered chromosome conformations. In fact, these not only maintain the correct preferential radial positioning of functional regions but further acquire a more specific organization in macro-domains that significantly overlaps with that recently identified by ref.~\cite{kalhor2012}.

Finally, we characterized the compliance of the optimized, steered chromosomes to reconfigure in a dense linear conformation. This test was aimed at mimicking the interphase$\to$mitotic rearrangements that occur during cell cycle, also aided by topoisomerases, without encountering significant topological hindrance.
The optimised chromosomes showed excellent compliance towards developing a dense linear organization. This
confirms that chromosome entanglement in the optimised system is minimal, and hence realistic.

To summarise, a consistent accord with phenomenological observations~\cite{federico2008,bickmore2013} was found for all considered properties of the optimised model chromosome configurations, from the preferred radial positioning of several types of functionally-relevant regions to the capability of chromosomes to recondense and segregate without topological hindrance. It is notable that these features emerge by using Hi-C data\cite{dixon2012} and radial placement\cite{Bolzer2005} as the sole source of constraints for the general, physics-based chromosome models. This highlights the significant extent to which functionally-related aspects of the genomic organization principles can be extracted from experimental structural data with the aid of suitable data analysis and chromosome modelling. As a further proof of that, we recall that none of the preferential positioning properties of the monitored functional \emph{loci} emerge without imposing any phenomenological constraints.

We expect that the general approach followed in this study could be profitably extended and transferred to other systems, so to incorporate additional knowledge-based information and to  capture more detailed aspects of the spatial organizational principles of eukaryotic nuclei. These will be even more relevant when missing genomic repeat structures eventually become incorporated into analysis. Furthermore, we in particular envisage that extending considerations from {\em cis}- to {\em trans}-chromosome contacts may be important towards pinning down more precisely the relative positioning of chromosomes, and also their absolute position in the nuclear environment.

\section*{Methods}
\noindent {\bf Significant pairwise constraints from Hi-C data.} The phenomenological constraints were based on two sources: primarily the raw data of ref.~\cite{dixon2012} for human lung fibroblasts (IMR90) and embryonic stem cells (hESC) and the list of significant contacts provided by Rao \emph{et al.} based on their high resolution experiments on IMR90~\cite{rao2014}.

The significant IMR90 and hESC pairings from the data of ref.~\cite{dixon2012} were singled out as follows. As input data, $x_{ij}$, we took the number of Hi-C reads between two chromosome regions, $i$ and $j$, discretized at the $100$kb resolution (comparable to the coarse-graining level of the chromosome model) and corrected for technical bias with the method of Imakaev \emph{et al.}~\cite{imakaev2012}. The distribution of $x$ values at fixed genomic distance, $\delta \equiv | i - j| $ was modelled using the discrete zero-inflated negative binomial mixture distribution (ZiNB)~\cite{yau2003}.
Specifically, the distribution of all reads with a given $\delta$ was fitted using the following probability distribution function:
\begin{align}
  P(X = 0 | \theta, p, \pi) &= \pi + (1- \pi)p^{\theta}  \\
  P(X = x | \theta, p, \pi) &= (1- \pi) NB(X = x | \theta, p ) \notag \\
  &= (1- \pi) \frac{\Gamma(x+\theta)}{\Gamma(\theta)x!}p^{\theta}(1-p)^x \, \, \text{ for } x = 1, 2, 3, \ldots, \max(\delta)
\end{align}
\noindent In the above expressions, $p$ is the probability of not having a 3D contact, $\theta$ captures the extra variance of the data with respect to a Poisson distribution, and $\pi$ is the probability of observing additional zeros in the data set. These reflect the intrinsic sparsity of Hi-C data sets, which cannot be accounted for by the sole negative binomial distribution.
The fitting procedures were carried out with the \emph{pscl} package in R~\cite{zeileis2007}.

Based on this ZiNB model distribution, the $p$-value for each $x_{ij}$, that is the probability of observing a contact frequency equal to or more extreme than $x_{ij}$, is given by:
\begin{align}
  p_{value}(x_{ij}) &= P_{\delta}(X \geq x_{ij} | \hat{\theta}, \hat{p}, \hat{\pi} ) \notag \\
  &=  (1- \hat{\pi} ) NB(X \geq x_{ij} | \hat{\theta}, \hat{p})
\end{align}
\noindent where $\hat{\theta}$, $\hat{p}$, and $\hat{\pi}$ are the ($\delta$-dependent) best-fit parameters.
We correct for multiple testing at fixed $\delta$ by selecting significant interactions at 1\% false-discovery rate of the Benjamini-Hochberg method~\cite{benjamini1995}. The criterion yields a total of $16,409$ and $14,928$ interactions (respectively $0.07$\% and $0.06$\% of all possible \emph{cis-}chromosome pairs between $100$kb regions), for lung fibroblasts (IMR90) and for embryonic stem cells (hESC) data sets respectively. The significant contacts for IMR90 are listed in Supplementary Table S1, see Supplementary Fig. S$1$ for some of their graphical representations, and those of hESC are provided in Supplementary Table S$2$.

We note that the multiple testing correction assumes that Hi-C maps entries are uncorrelated. Even accounting for the medium resolution of the data sets, this assumption can hold only approximately. Appropriate sampling strategies have been suggested for dealing with such correlations in ref.~\cite{rao2014}. While these strategies are not primed to be used in conjunction with our analysis, the viability of our method is shown {\em a posteriori}. In fact, a similar compliance to steering is observed for the contacts selected using the data of ref.~\cite{dixon2012} and our statistical method, and for those selected using the data and the statistical analysis of ref.~\cite{rao2014}. Furthermore, we emphasize that accounting for correlations during multiple testing correction can only result in a less conservative correction procedure, and will not reduce the number of false positives. Thus, in this analysis, the effect would be minor.

{\bf Genome-wide modelling and spatial constraints.} The statistically significant Hi-C contacts were enforced as spatial constraints for a genome-wide modelling of chromosome organization in lung fibroblasts (IMR90) and embryonic stem cells (hESC) nuclei.

The model chromosomes are packed at the typical nuclear density of $0.012$ bp/nm$^3$ inside a confining nucleus. For simplicity, the nuclear shape has been chosen to be spherical with a radius of $4,800$ nm, neglecting the flattened ellipsoidal shape of fibroblast cells~\cite{Bolzer2005}. Each chromosome is treated as a chain of beads~\cite{kremer_jcp} with diameter $\sigma=30$ nm equal to the nominal chromatin fiber thickness, and hence corresponding to a stretch of $3.03$ kb~\cite{plospaper,bjpaper,distefano2013}. The chain bending rigidity is chosen so that the model chromatin fiber has the correct persistence length ($150$ nm)~\cite{plospaper}.

The chromosome lengths, in number of beads, are given in Supplementary Table S$3$ and range from $15,873$ to $82,269$ (for chromosome $21$ and $1$, respectively). The total number of beads in the system is $1,952,709$, resulting from the presence of two copies for each autosome and a single copy of the sexual chromosome X. The latter is present in a single copy to account for the absent (male) or inactivated (female) copy, while the male chromosome Y is not considered for its small size. Each chromosome is initially prepared in a rod-like structure resulting from stacked rosette patterns, as in ref.~\cite{plospaper}.

The rod-like chromosomes are initially positioned in the nucleus with two different protocols: random and phenomenological, that is matching the preferential chromosome radial positioning reported in ref.~\cite{Bolzer2005}. In the random scheme, the chromosomes are consecutively positioned, from the longest to the shortest, by placing their midpoint randomly inside the nucleus and with a random axis orientation. If \emph{trans} steric clashes arise at any stage, the placement procedure is repeated. The phenomenological scheme is analogous to the random one, except that chromosome midpoints are placed within one of six discretised radial shells, so as to match as close as possible (best positioning out of $10,000$ trials) to the experimental average distance from the nuclear center reported in ref.~\cite{Bolzer2005}. Chromosome protrusions beyond the nuclear sphere are next eliminated by briefly evolving the system with a stochastic (Langevin) molecular dynamics simulation under the action of a radial compressive force, until all beads are brought inside the spherical nuclear boundary, which is treated as impenetrable for all simulations. The Langevin dynamics is integrated numerically with the LAMMPS simulation package~\cite{lammps}, with default parameters and the duration of this initial compressive adjustment phase is set equal to $1,200$ $\tau_{LJ}$, where $\tau_{LJ}$ is the characteristic Lennard-Jones simulation time (see Supplementary Methods).

The chromosomes are next evolved with the stochastic dynamics for a timespan of $120,000$ $\tau_{LJ}$ (corresponding to $7$ hours in realtime using the time mapping in ref.~\cite{plospaper}) during which centromeres are compactified by the attractive pairwise interaction of their constitutive beads while chromosome arms relax from the initial rod-like and expand inside the nucleus, adapting to its shape (see Supplementary Methods).
After this relaxation phase, the system dynamics is steered to promote the formation of target \emph{cis}-chromosome pairings corresponding to the statistically significant Hi-C contacts. As in ref.~\cite{distefano2013}, the steering involves setting harmonic constraints between the centers of mass of paired target regions, each spanning 33 beads (equivalent to the $100$kb data resolution). To minimize the out-of-equilibrium driving of the system, the strength of the spring constant is progressively ramped up during the steering phase, which lasts for a maximum of $6,000$ $\tau_{LJ}$, starting from suitably weak initial values of the spring constants. The latter are set based on the sequence separation of the target pair so as to just counteract their entropic recoil (see Supplementary Methods). The same steering procedure has been applied to the optimal conformations (only for IMR90 cell line) to enforce the contacts of ref.~\cite{rao2014} for a maximum of $600$ $\tau_{LJ}$. All the steered simulations have been performed using the PLUMED~\cite{plumed} package for LAMMPS.

Both the system relaxation and the steering procedure are replicated independently $10$ times per each of the $4$ considered setups, in order to assess the average properties of the models. The four setups are given by the combination of the IMR90 or hESC cell lines and the phenomenological or random chromosome positioning. To ensure the statistical independence, we used one conformation per run, namely the snapshot taken at the end of the steering protocol, for the analysis of the structurally and functionally-oriented properties of the nuclear chromosome organization.\\
The cumulative computer time of the $40$ simulations involved $150,000$ single CPU hours on $32$ processors for the pre-steering runs, and $15,000$ single CPU hours on a single processor for the steered ones.

{\bf Nuclear positioning of functionally-related {\em loci}.} The steered genome-wide organization was profiled for the preferred positioning of various functionally-related {\em loci}: genes, lamina-associated domains (LADs), H3K4me3 (activating), H3K9me3 and H3K27me3 (repressive) histone modifications, and Giemsa staining bands. The genomic location of genes and bands were obtained from the UCSC Table Browser~\cite{karolchik2004}, those of LADs from ref.~\cite{guelen2008}, and histone modification data were obtained from GEO accession number $GSE16256$. These regions were mapped on the model chromosomes and their preferential radial positioning, before and after steering, was characterized by subdividing the nucleus in $15$ radial shells of equal thickness ($\sim 320$ nm), and computing for each shell the enrichment in beads associated to a given functional region.

{\bf Spatial macrodomains.} The macrodomains of chromosome 19, which has the highest density of imposed constraints ($\sim 15$ constraints/Mb, see Supplementary Table S$3$), were obtained from a spatial clustering analysis and then compared with the ``block'' partitions established in ref.~\cite{kalhor2012} based on tethered conformational capture techniques. The clustering consisted of a sequence-continuous K-medoids partitioning of chromosome arms, discretised at the 100kb-long level, corresponding to $33$ beads. The entries of pairwise dissimilarity matrices, $\Delta$, were based on the average distances, $\langle d \rangle$,  of the 100kb-long segments, the average being taken over the final conformations optimized with the constraints of refs.~\cite{dixon2012,rao2014}. Specifically, for two segments $i$ and $j$, the corresponding dissimilarity entry, $\Delta_{ij}$ was set equal to $\langle d_{ij} \rangle$ when the latter was below the $750$nm. The cutoff distance of $1,000$nm was used for more distance pairs. The two chromosome arms were separately subdivided in the same number of domains of ref.~\cite{kalhor2012} and the significance of the resulting correspondence, or overlap, of the sequentially numbered domains was obtained by comparison against $1,000$ random sequence-continuous subdivisions of the arms in the same number of domains.

{\bf The chromosome pre-mitotic recondensation.} To reconfigure the model chromosomes to a linear (mitotic-like) state we switched off the target harmonic constraints and replaced them with couplings between pairs of {\em loci} at the regular sequence separation of $200$ kb. These constraints promote the chromosome rearrangement into a linear succession of loops analogous to the mitotic chromosome models of refs.~\cite{sikorav1994,naumova2013}. During this simulation, the parameters of all the harmonic constraints are treated on equal footing. The spring constants are maintained equal and unvaried, while the equilibrium distances are decreased in steps of $30$ nm, from $200$ nm (the maximum extension of a $200$ kb chromatin model strand) to $30$ nm (the size of a bead), every $0.6$ $\tau_{LJ}$. At the final nominal equilibrium distance, the simulations are extended up to $300$ $\tau_{LJ}$. Besides applying the recondesation procedure to the $10$ steered replicates of the lung fibroblast (IMR90), we also apply it to the relaxed, pre-steering chromosomes.

\section*{Acknowledgements}
We are grateful to Marc A. Marti-Renom, Davide Ba\`u, and Angelo Rosa
for useful discussions. This work was supported by The Norwegian
Cancer Society [grant 71220 - PR-2006-0433]; the Research Council of
Norway; and the Italian Ministry of Education [grant PRIN 2010HXAW77].

\section*{Author contributions statement}
MDS JP EH CM conceived and designed the experiments; JP and TGL
performed statistical analyses; MDS performed the molecular dynamics
simulations; MDS and JP analysed the data and implemented the analysis
tools; MDS JP EH CM wrote the paper.

\section*{Additional information}
\textbf{Competing financial interests} The authors declare no competing financial interests..

\clearpage

\renewcommand{\figurename}{{\bf Supplementary Figure S}}
\setcounter{figure}{0}
\section*{Supplementary Figures}


\begin{figure}[h!]
  \centering
  \includegraphics[width=\textwidth]{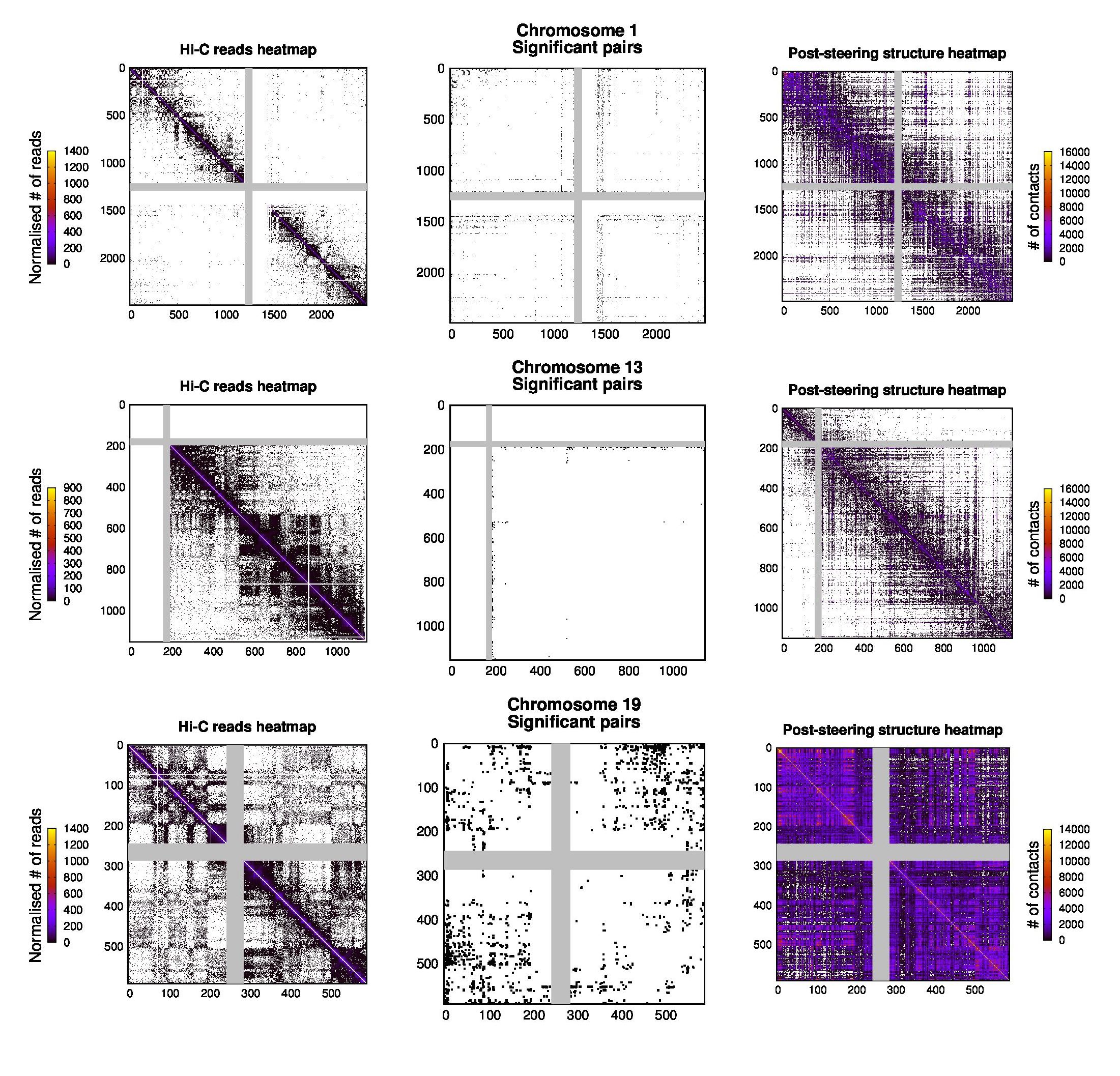}
\end{figure}
\begin{figure}
  \centering
  \includegraphics[width=\textwidth]{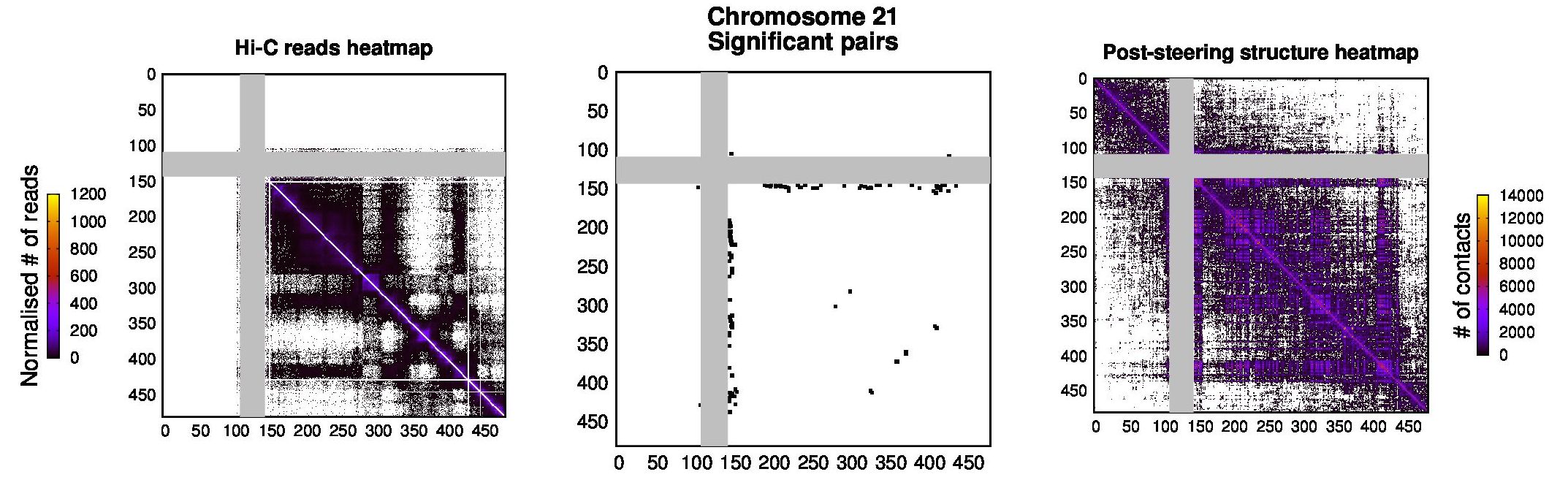}
  \caption{{\bf Cis-chromosome heatmaps for lung fibroblasts (IMR90) cell line at 100 kb resolution.} The panels on the left side show the Hi-C contact propensity maps of human embryonic stem cells from Dixon {\em et al.} (2012) that have been adjusted for technical biases using the method described in Imakaev {\em et al.} (2012). Central panels show the subset of statistically significant Hi-C contacts. The dots used for the entries have been magnified for visual clarity. Panels on the right side show the contact map (cutoff $240$nm) obtained from the $10$ optimally-steered models of IMR90 nuclei with phenomenological initial positioning from Bolzer {\em et al.} (2005). Only selected chromosomes are represented, namely: chr1 which is the longest, chr21 which is the shortest, chr19 which has the largest sequence-wise density of target constraints, and chr13 which has the smallest one. In all panels, the gray bands mark the centromeric region}
\end{figure}

\begin{figure}[h!]
  \centering
  \includegraphics[width=\textwidth]{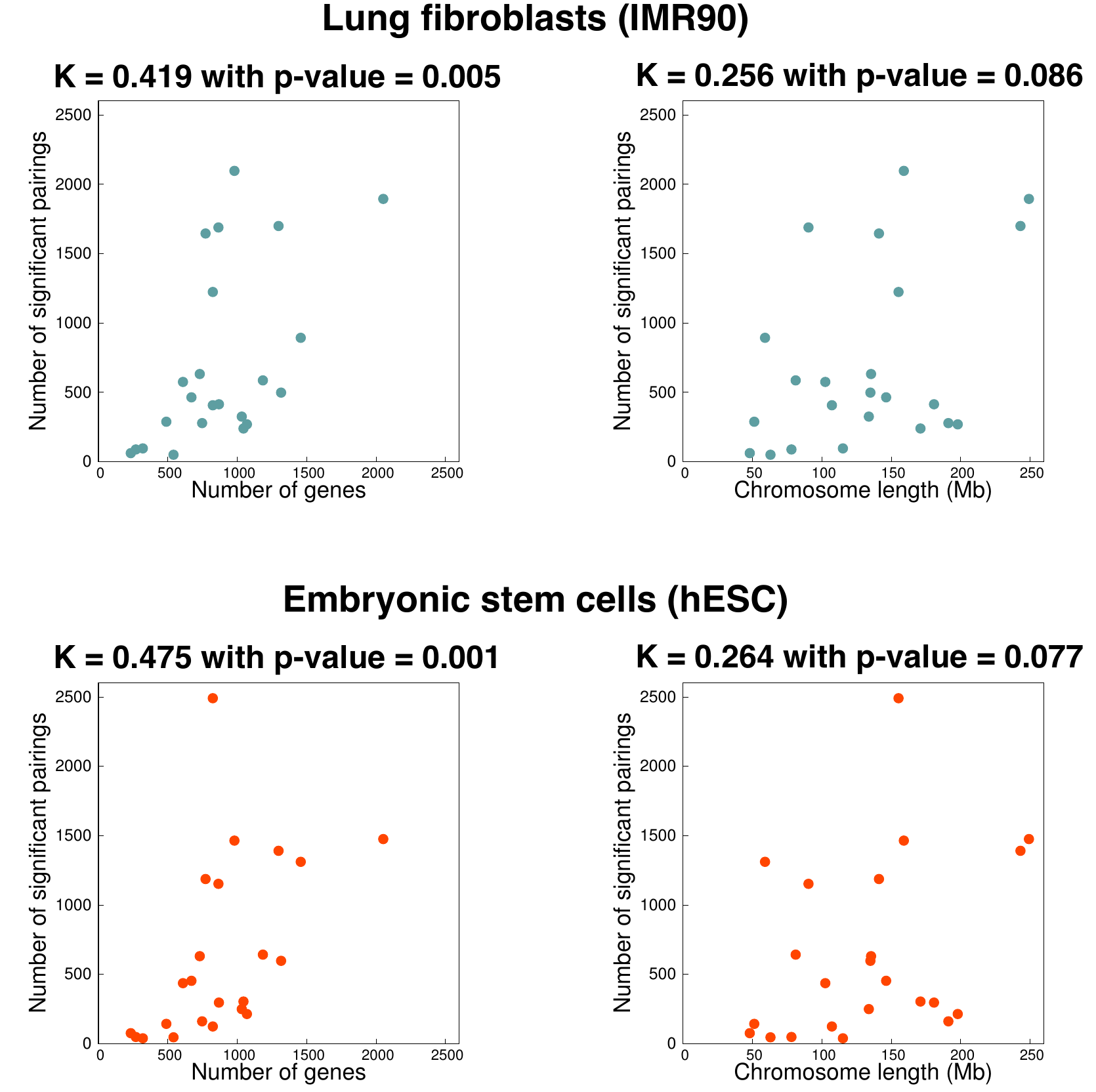}
  \caption{{\bf Number of significant pairings per chromosome based on the analysis of the data in Dixon \emph{et al} (2012) for lung fibroblasts (IMR90) and embryonic stem cells (hESC).} For both cell lines, the number of significant pairings correlates significantly with the number of genes in the chromosomes ($p-value<0.005$ of non-parametric Kendall rank-correlation), but only weakly with chromosome length ($p-value>0.08$).}
\end{figure}


\begin{figure}[ht!]
  \centering
  \includegraphics[width=\textwidth]{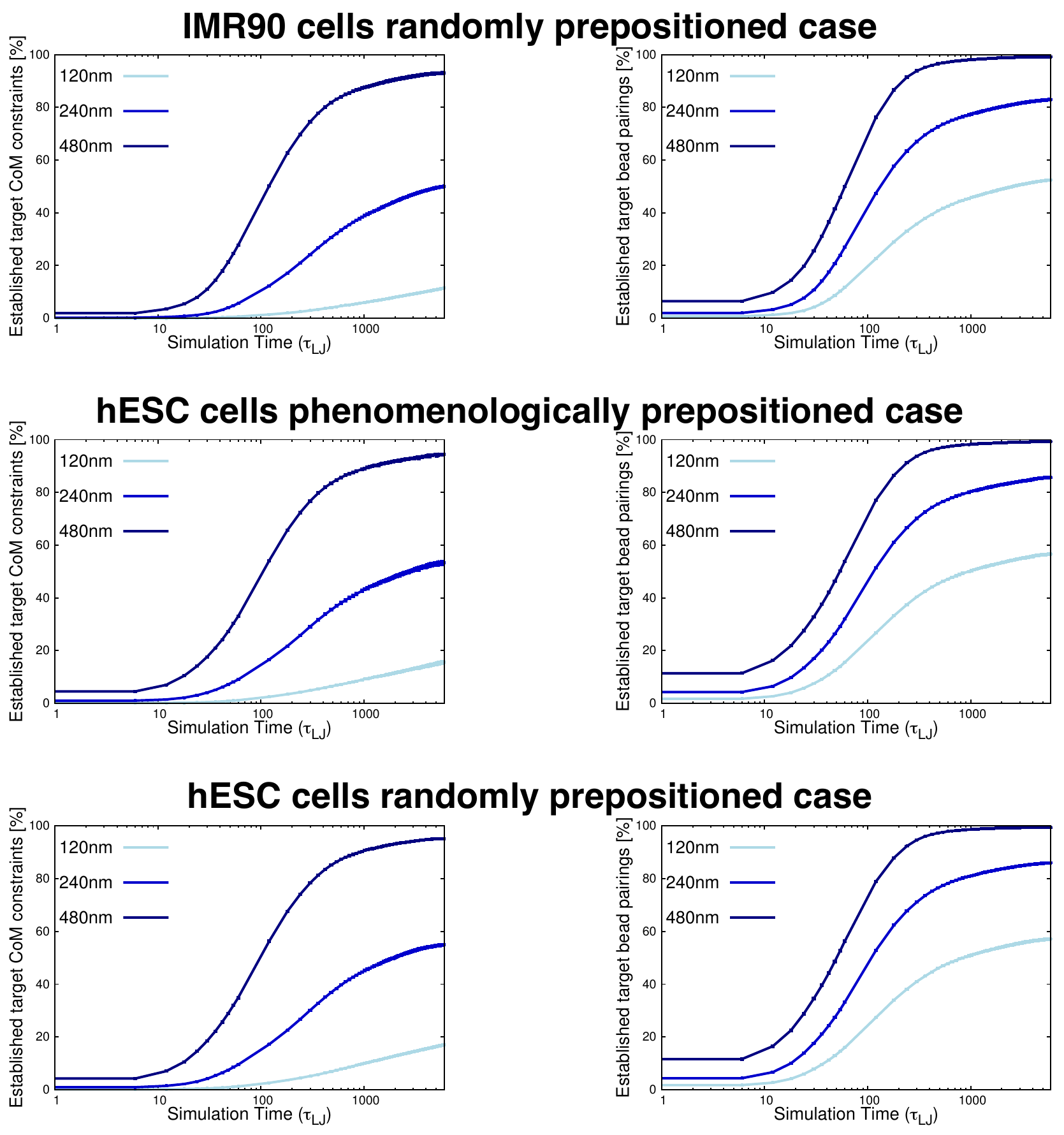}
  \caption{{\bf Evolution of the satisfied target constraints from the analysis of the data in Dixon \emph{et al} (2012) for different cell lines and chromosome positioning schemes}. The curves show the increase of the percentage of target contacts that are established in the course of the steering dynamics for lung fibroblasts cells (IMR90) and human embryonic stem cells (hESC) nuclei and for different chromosome positioning schemes (phenomenological and random). The plots complement the information provided in Fig. $1$ of the main text.}
\end{figure}


\begin{figure}[!ht]
  \centering
  \includegraphics[width=\linewidth]{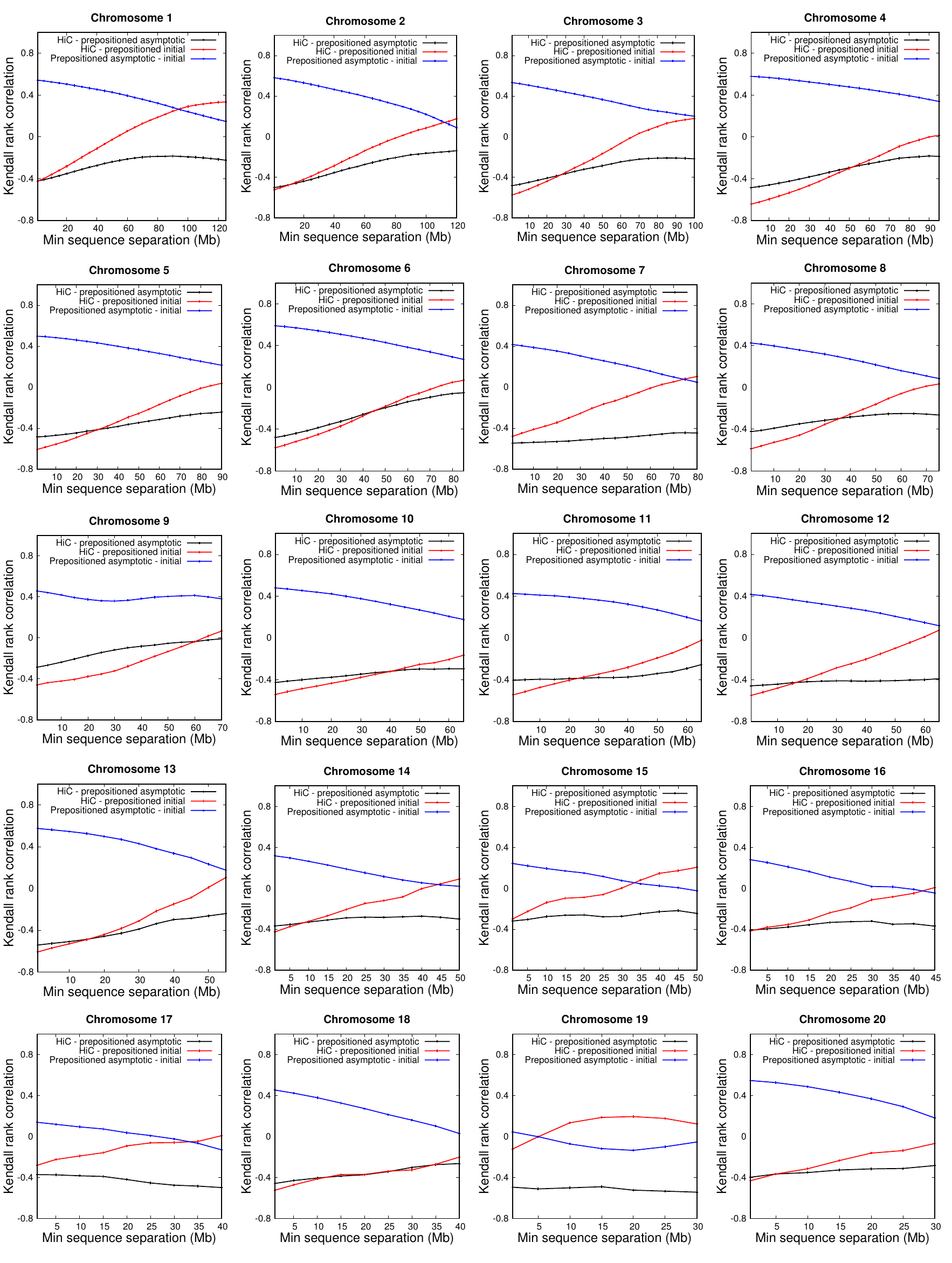}
\end{figure}
\clearpage
\begin{figure}[!ht]
  \centering
  \includegraphics[width=\linewidth]{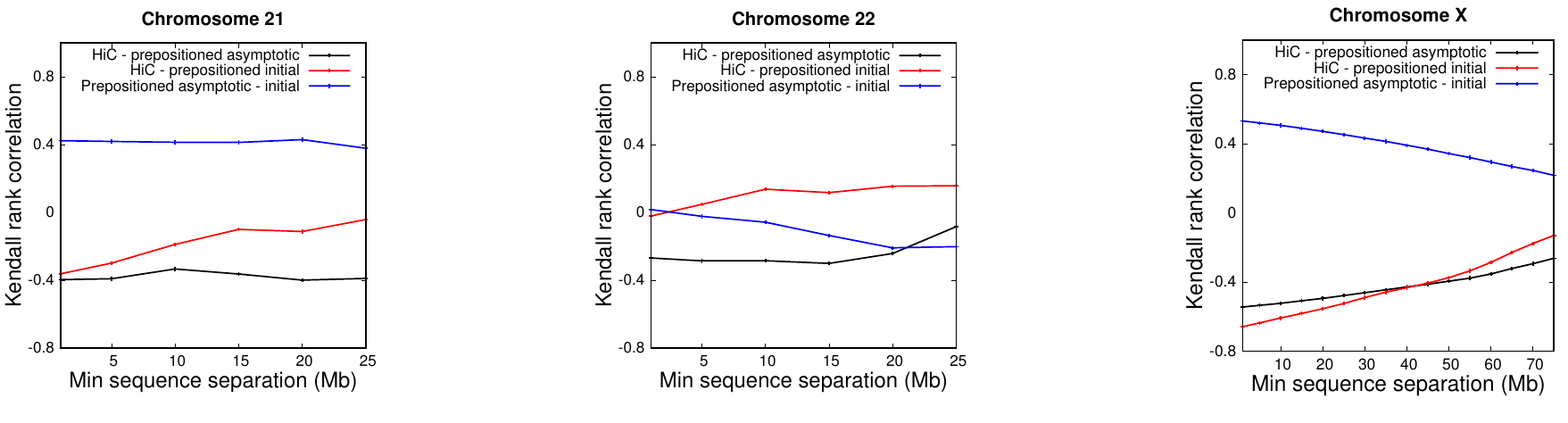}
  \caption{{\bf Correlation analysis between the Hi-C contact matrices
      and the models' distance matrices.}. The curves show the {\em
      cis-}chromosome Kendall rank correlation coefficients between
    different matrix pairs considering entries at different minimum
    sequence separations from $1$Mb up to half of the chromosome
    length. The entries of each {\em cis-}chromsome Hi-C contact
    matrix from Dixon {\em et al.} (2012) are compared with the
    entries of each model distance matrix in the prepositioned case
    after steering (black curves) and in the prepositioned case before
    steering (red curves). For completeness, the two model distance
    matrices are also compared between them (blue curves). Due to the
    high number of entries, Kendall correlation coefficients larger
    than 0.147 in modulus are statistically significant because they
    have a (two-sided) $p-value$ smaller than 0.05 for the smallest
    chromosomes, and hence a much lower p-value for the other, longer
    ones.}
\end{figure}
\clearpage

\begin{figure}[ht!]
  {\bf Lung fibroblasts (IMR90) nuclei with phenomenological prepositioning of the chromosomes}\\
  \centering
(a) \includegraphics[width=0.4\textwidth]{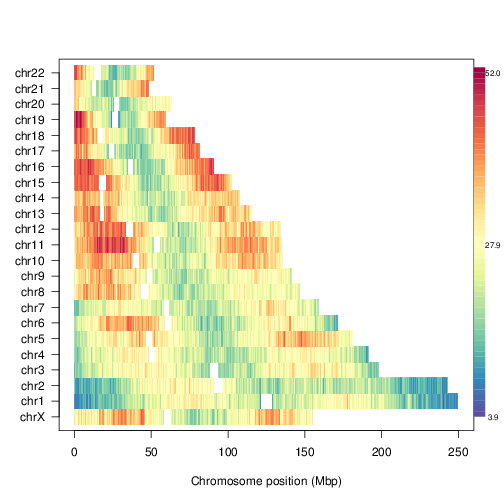}
(b) \includegraphics[width=0.4\textwidth]{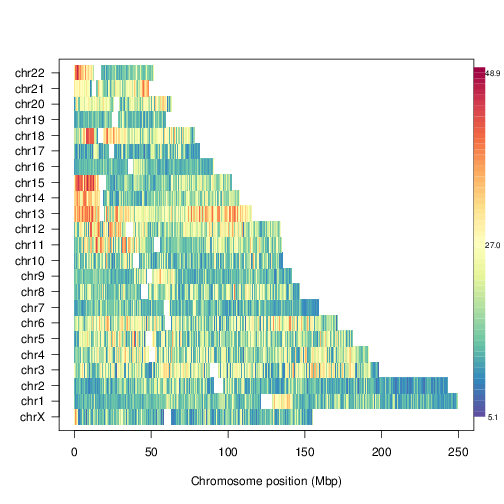}
  \caption{{\bf Genome-wide variability of radial bead position.} The panels are based on configurations obtained starting from the phenomenological prepositioning of the chromosomes in Bolzer {\em et al.} (2005), immediately prior to (a) and after (b) applying the steering protocol appropriate for IMR90 cell line based on the analysis of the data in Dixon \emph{et al} (2012). Numbers indicate the standard deviation of the radial position across the $10$ replicate simulations. The plots complement the information provided in Fig. $3$ of the main text.}
\end{figure}

\begin{figure}[ht!]
  {\bf Lung fibroblasts (IMR90) nuclei with random prepositioning of the chromosomes}\\
  \centering
(a)  \includegraphics[width=0.4\textwidth]{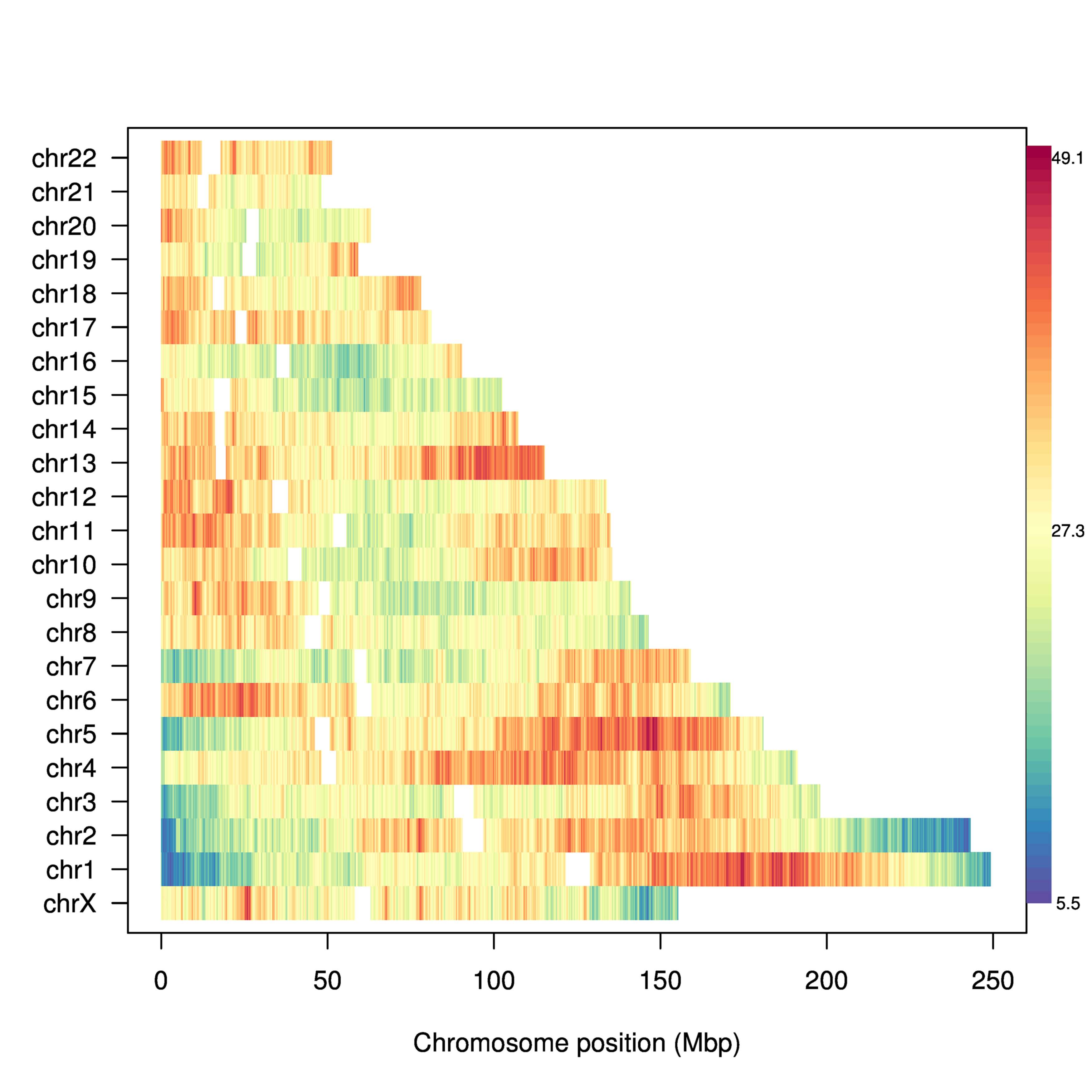}
(b)  \includegraphics[width=0.4\textwidth]{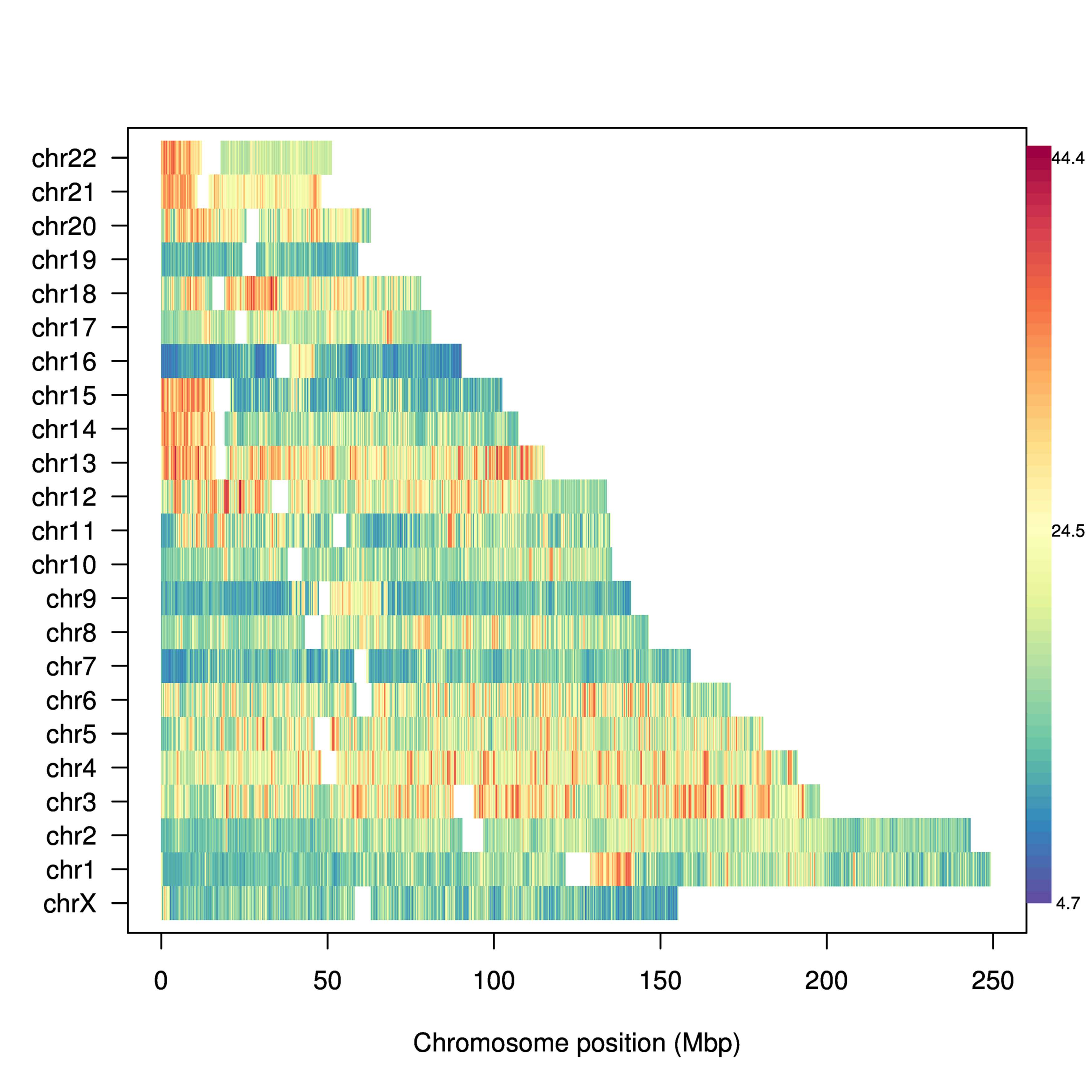}
  \caption{{\bf Genome-wide variability of radial bead position.} The panels are based on configurations obtained starting from a random prepositioning of the chromosomes, immediately prior to (a) and after (b) applying the steering protocol appropriate for IMR90 cell line based on the analysis of the data in Dixon \emph{et al} (2012). Numbers indicate the standard deviation of the radial position across the $10$ replicate simulations.}
\end{figure}

\begin{figure}[ht!]
  {\bf Embryonic stem cells (hESC) nuclei with phenomenological prepositioning of the chromosomes}\\
  \centering
(a)  \includegraphics[width=0.4\textwidth]{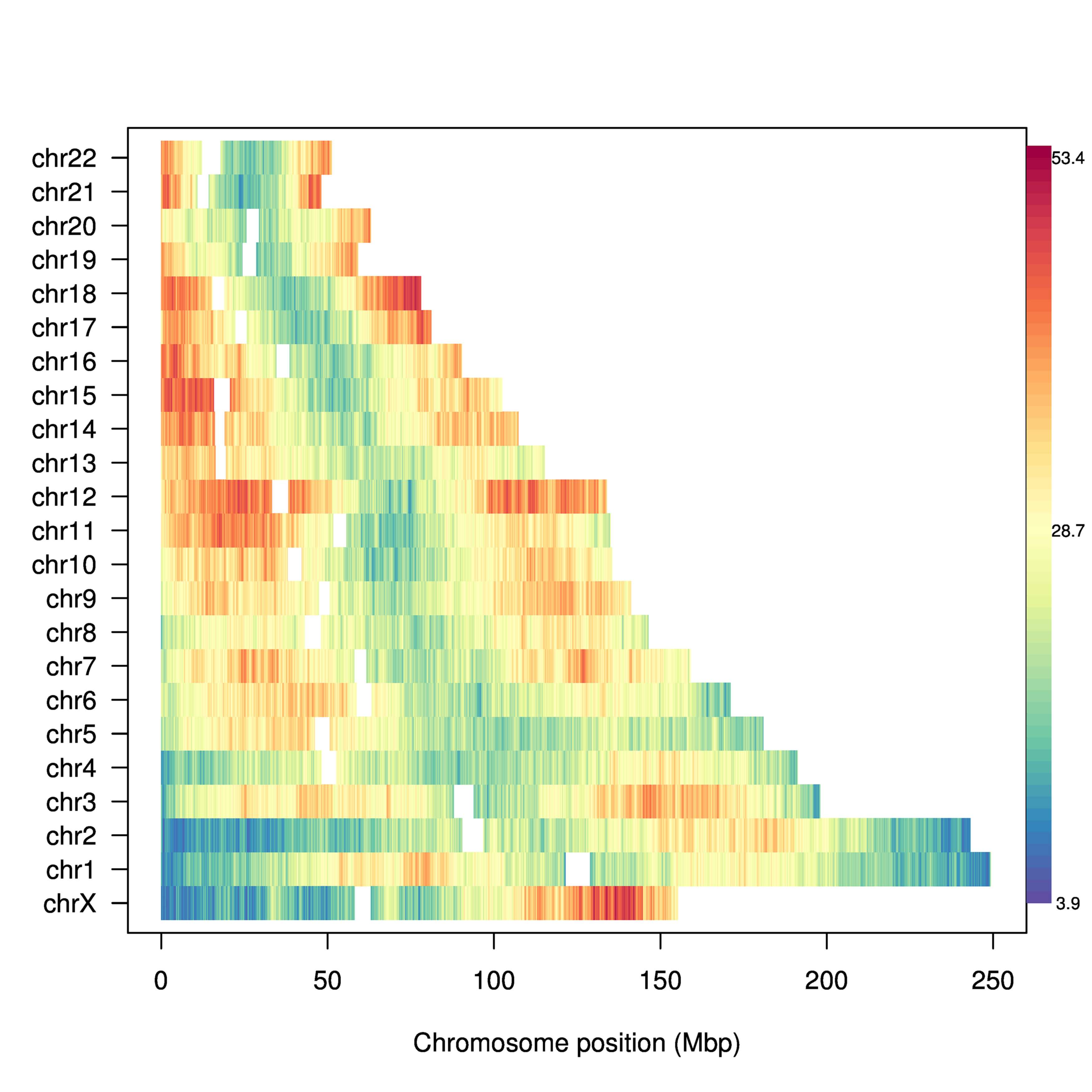}
(b)  \includegraphics[width=0.4\textwidth]{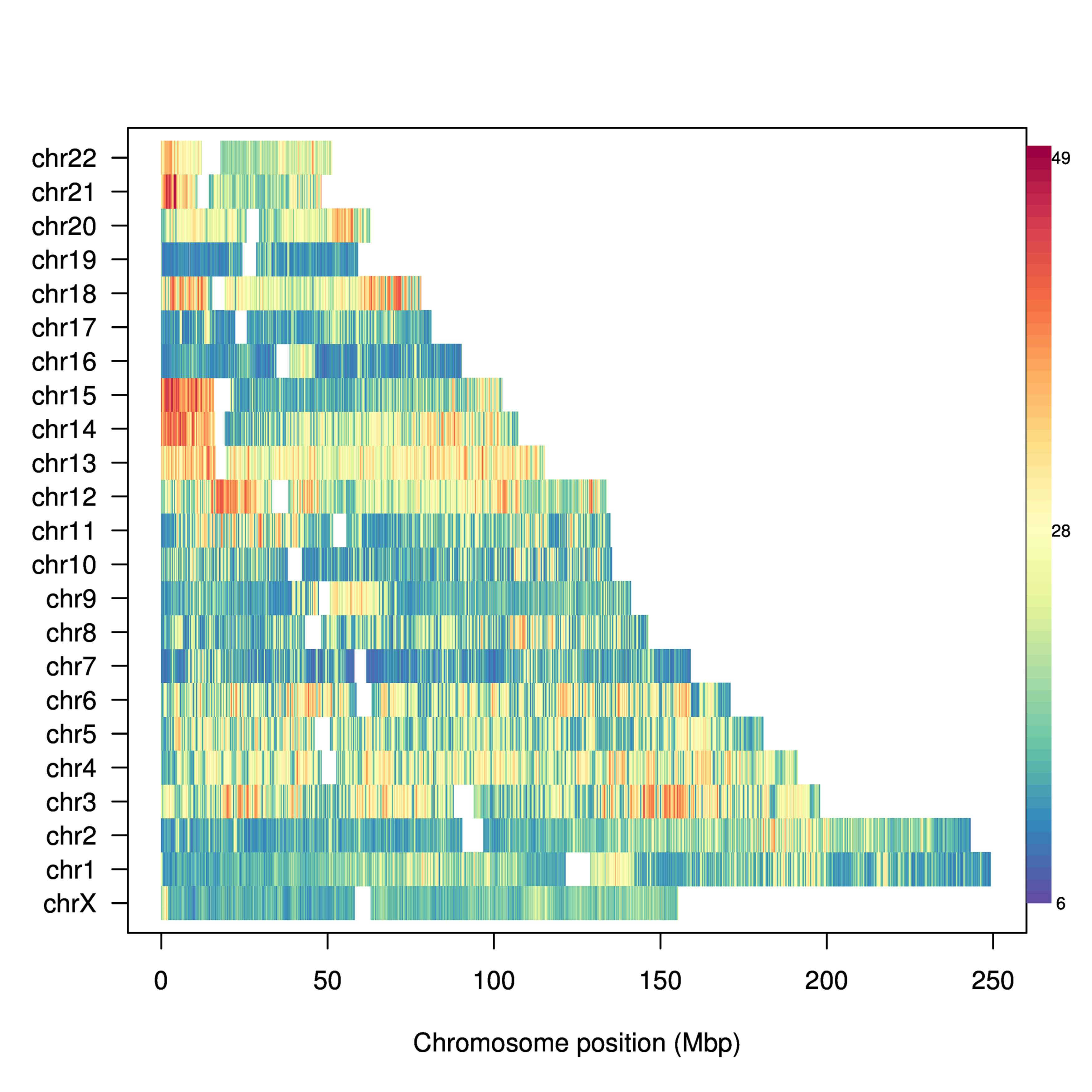}
  \caption{{\bf Genome-wide variability of radial bead position.} The panels are based on configurations obtained starting from the phenomenological prepositioning of the chromosomes in Bolzer {\em et al.} (2005), immediately prior to (a) and after (b) applying the steering protocol appropriate for hESC cell line based on the analysis of the data in Dixon \emph{et al} (2012). Numbers indicate the standard deviation of the radial position across the $10$ replicate simulations.}
\end{figure}

\begin{figure}[ht!]
  {\bf Embryonic stem cells (hESC) nuclei with random prepositioning of the chromosomes}\\
  \centering
(a)  \includegraphics[width=0.4\textwidth]{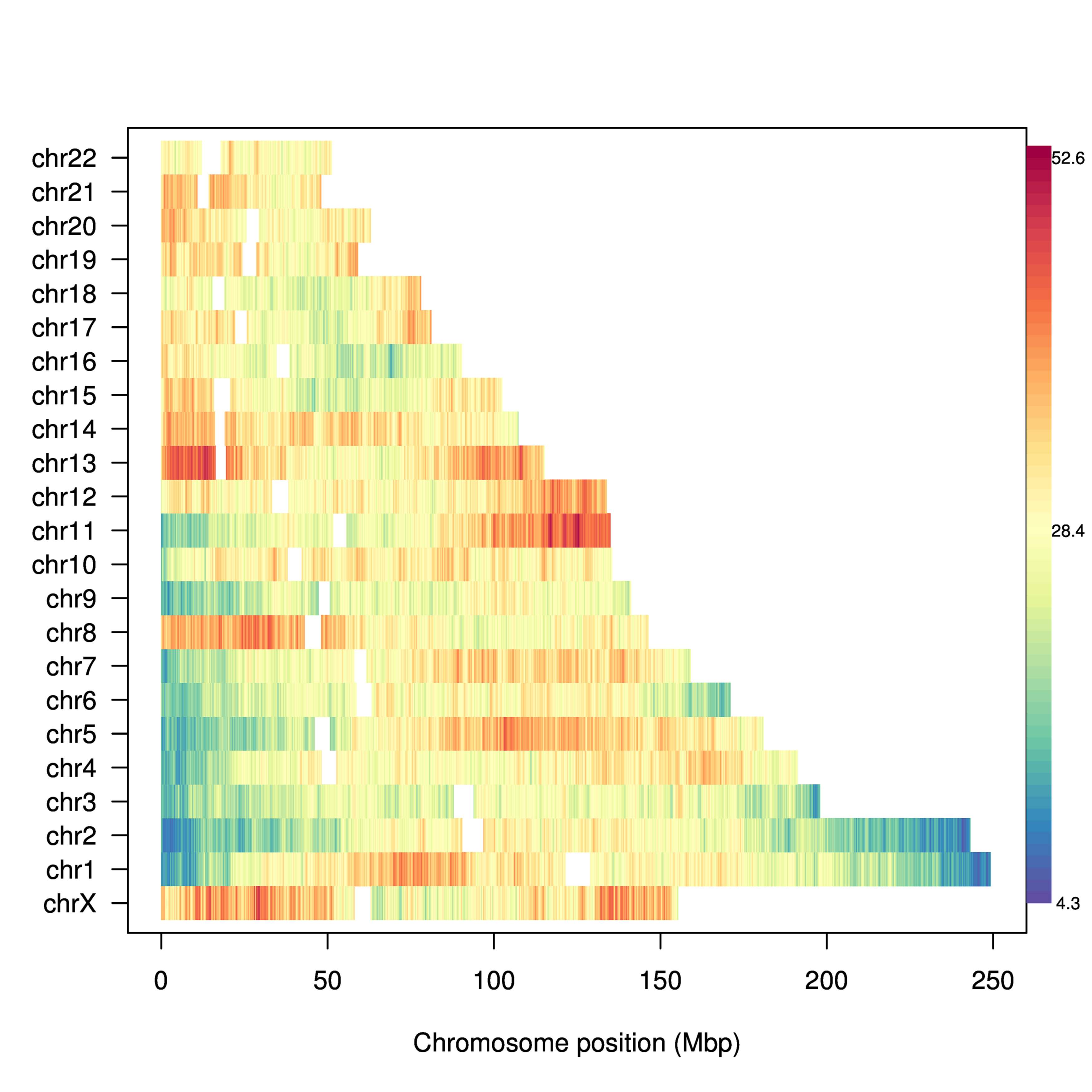}
(b)  \includegraphics[width=0.4\textwidth]{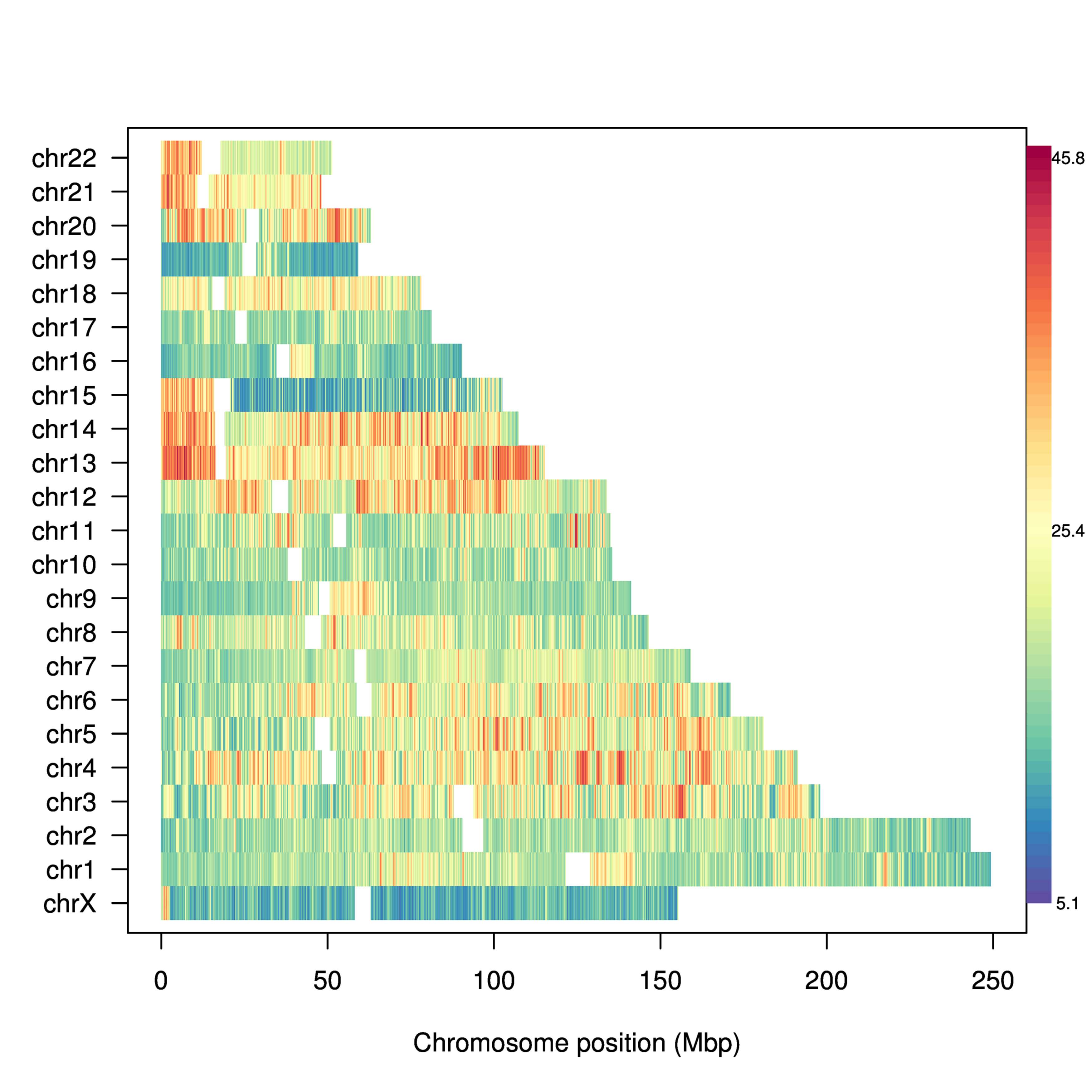}
  \caption{{\bf Genome-wide variability of radial bead position.} The panels are based on configurations obtained starting from a random prepositioning of the chromosomes, immediately prior to (a) and after (b) applying the steering protocol appropriate for hESC cell line based on the analysis of the data in Dixon \emph{et al} (2012). Numbers indicate the standard deviation of the radial position across the $10$ replicate simulations.}
\end{figure}


\begin{figure}[ht!]
  {\bf Lung fibroblasts (IMR90) nuclei with phenomenological prepositioning of the chromosomes}\\
  \centering
(a)  \includegraphics[width=0.4\textwidth]{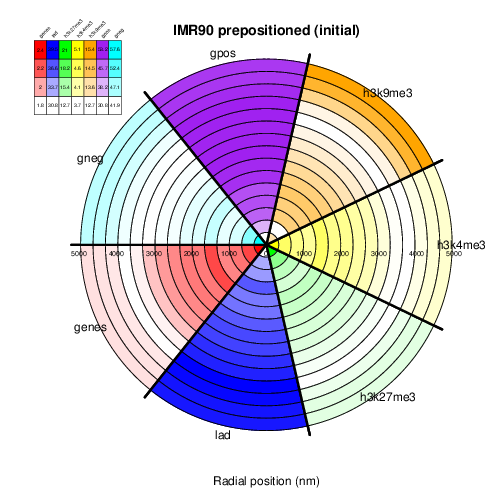}
(b)  \includegraphics[width=0.4\textwidth]{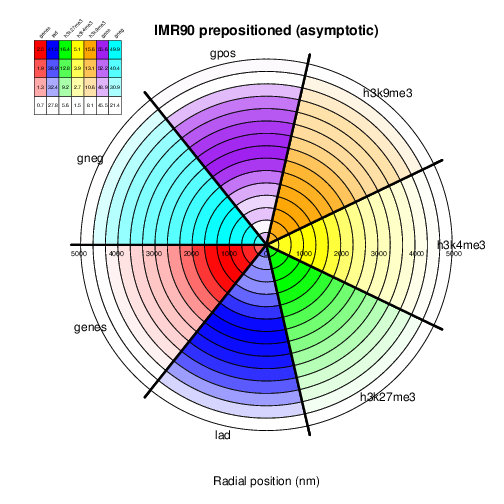}
  \caption{{\bf Nuclear positioning of functionally-oriented genomic regions.} Circular histograms giving the relative density (in percent) based on H3K9me3 (orange), H3K4me3 (yellow), H3K27me3 (green), LADs (blue) and genes (red), and negative (cyan) and positive (purple) Giemsa staining bands. Circle slices (thickness $\sim 320$nm) indicate radial position (in nm) within the nucleus aggregated across all $10$ replicate simulations, and the numbers indicate the percentage of beads associated with the given feature relative to the total number of beads in the given circle slice. The figure is based on configurations obtained starting from the phenomenological prepositioning of the chromosomes in Bolzer {\em et al.} (2005), immediately prior to (a) and after (b) applying the steering protocol appropriate for IMR90 cell line based on the analysis of the data in Dixon \emph{et al} (2012).}
\end{figure}

\begin{figure}[ht!]
  \centering
    \includegraphics[width=\textwidth]{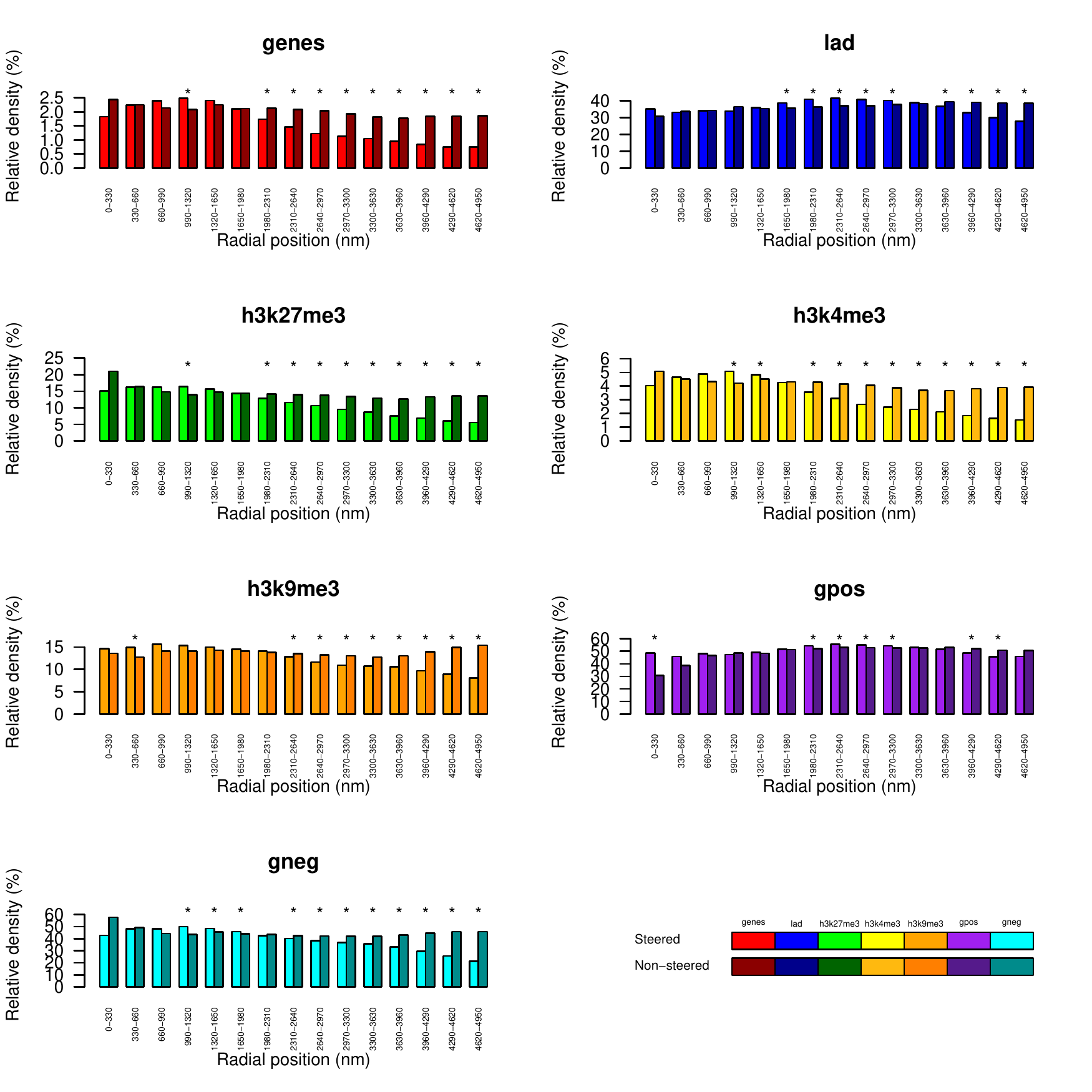}
    \caption{{\bf Histograms of the relative density of genes, LADs, H3K27me3, H3K4me3, H3K9me3, and positive (gpos) and negative (gneg) Giemsa staining bands in concentric equally thick radial shells of the nucleus.} The figure is based on configurations obtained starting from the phenomenological prepositioning of the chromosomes in Bolzer {\em et al.} (2005), immediately prior to (Non-steered) and after (Steered) applying the steering protocol appropriate for IMR90 cell line based on the analysis of the data in Dixon \emph{et al} (2012). The asterisks indicate statistically significant differences between the two cases, using the Wilcoxon test with a p-value cutoff of $0.05$. The plots complement the information provided in Fig. $4$ of the main text.}
\end{figure}

\begin{figure}[ht!]
  {\bf Lung fibroblasts (IMR90) nuclei with random prepositioning of the chromosomes}\\
  \centering
    (a) \includegraphics[scale=0.4]{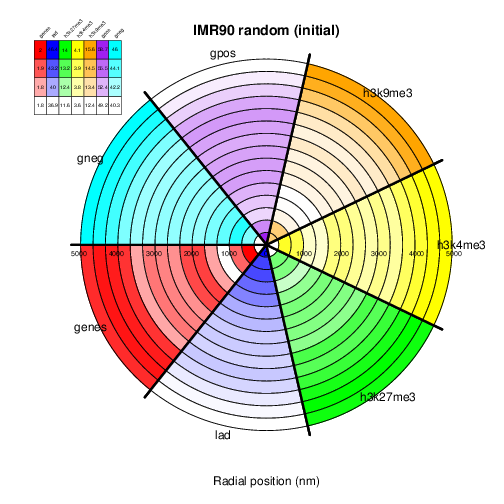}
    (b) \includegraphics[scale=0.4]{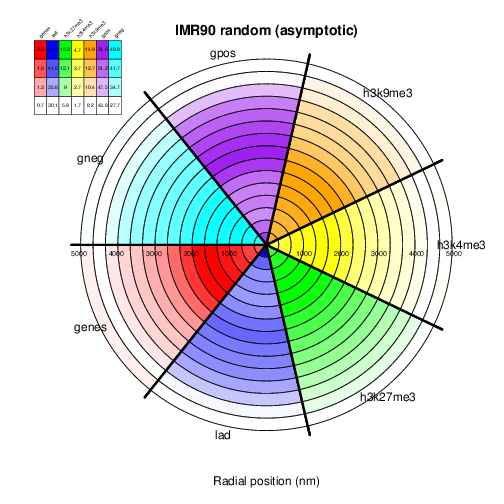}
  \caption{{\bf Nuclear positioning of functionally-oriented genomic regions.} Circular histograms giving the relative density (in percent) based on H3K9me3 (orange), H3K4me3 (yellow), H3K27me3 (green), LADs (blue) and genes (red), and negative (cyan) and positive (purple) Giemsa staining bands. Circle slices (thickness $\sim 320$nm) indicate radial position (in nm) within the nucleus aggregated across all $10$ replicate simulations, and the numbers indicate the percentage of beads associated with the given feature relative to the total number of beads in the given circle slice. The figure is based on configurations obtained starting from a random prepositioning of the chromosomes, immediately prior to (a) and after (b) applying the steering protocol appropriate for IMR90 cell line based on the analysis of the data in Dixon \emph{et al} (2012).}
\end{figure}

\begin{figure}[ht!]
  \centering
    \includegraphics[width=\textwidth]{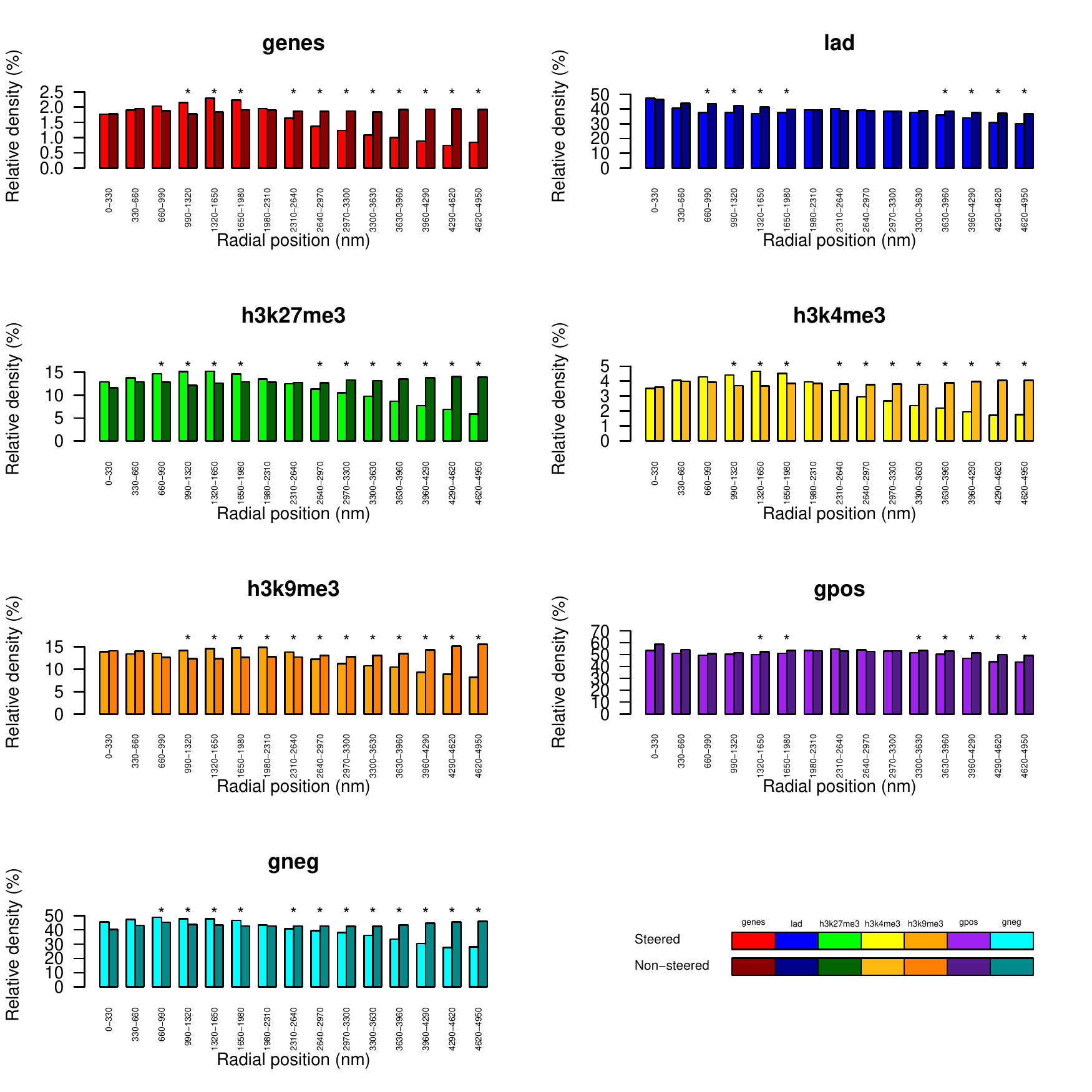}
    \caption{{\bf Histograms of the relative density of genes, LADs, H3K27me3, H3K4me3, H3K9me3, and positive (gpos) and negative (gneg) Giemsa staining bands in concentric equally thick radial shells of the nucleus.} The figure is based on configurations obtained starting from a random prepositioning of the chromosomes, immediately prior to (Non-steered) and after (Steered) applying the steering protocol appropriate for IMR90 cell line based on the analysis of the data in Dixon \emph{et al} (2012).}
\end{figure}

\begin{figure}[ht!]
  {\bf Embryonic stem cells (hESC) nuclei with phenomenological prepositioning of the chromosomes}\\
  \centering
  (a) \includegraphics[scale=0.4]{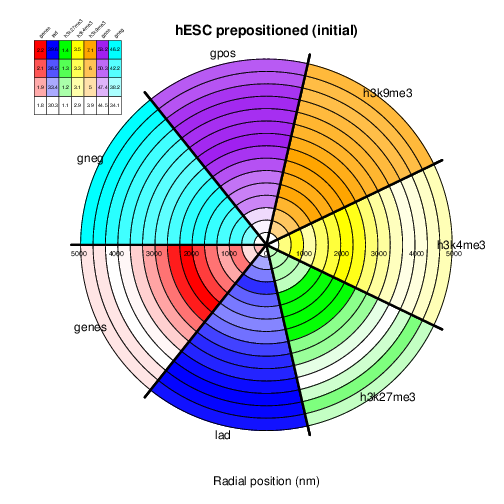}
  (b) \includegraphics[scale=0.4]{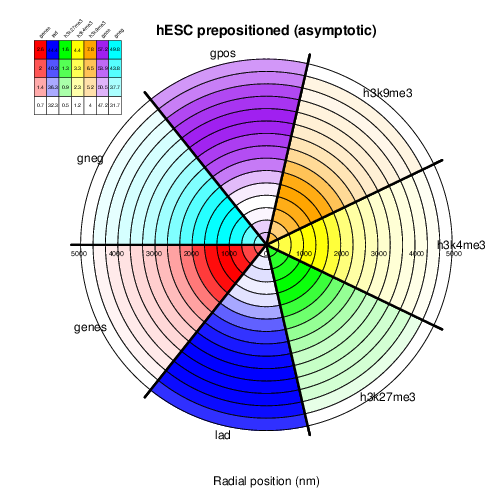}
  \caption{{\bf Nuclear positioning of functionally-oriented genomic regions.} Circular histogram giving the relative density (in percent) based on H3K9me3 (orange), H3K4me3 (yellow), H3K27me3 (green), LADs (blue) and genes (red), and negative (cyan) and positive (purple) Giemsa staining bands. Circle slices (thickness $\sim 320$nm) indicate radial position (in nm) within the nucleus aggregated across all $10$ replicate simulations, and the numbers indicate the percentage of beads associated with the given feature relative to the total number of beads in the given circle slice. The figure is based on configurations obtained starting from the phenomenological prepositioning of the chromosomes in Bolzer {\em et al.} (2005), immediately prior to (a) and after (b) applying the steering protocol appropriate for hESC cell line based on the analysis of the data in Dixon \emph{et al} (2012).}
\end{figure}

\begin{figure}[ht!]
  \centering
  \includegraphics[width=\textwidth]{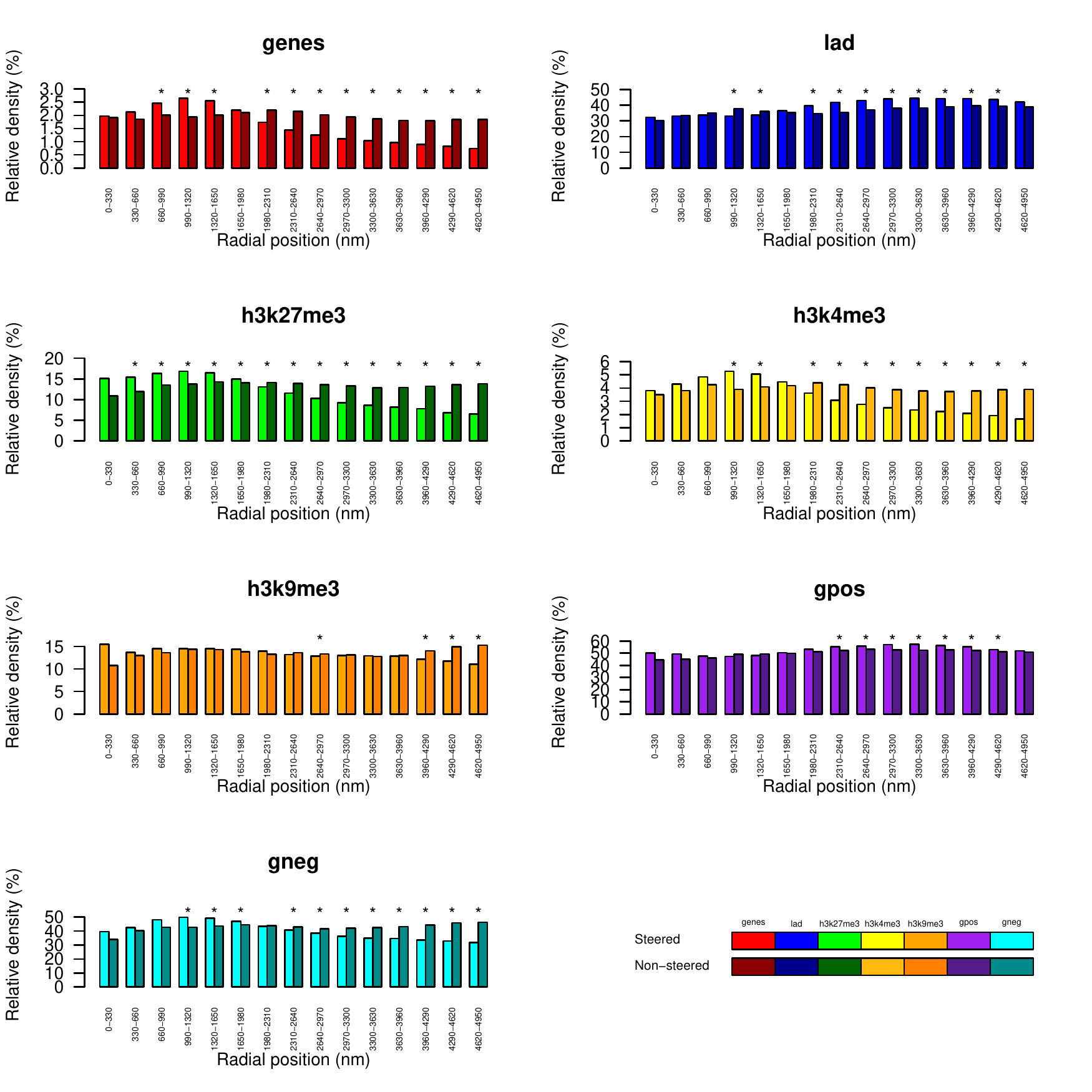}
  \caption{{\bf Histograms of the relative density of genes, LADs, H3K27me3, H3K4me3, H3K9me3, and positive (gpos) and negative (gneg) Giemsa staining bands in concentric equally thick radial shells of the nucleus.} The figure is based on configurations obtained starting from the phenomenological prepositioning of the chromosomes in Bolzer {\em et al.} (2005), immediately prior to (Non-steered) and after (Steered) applying the steering protocol appropriate for hESC cell line based on the analysis of the data in Dixon \emph{et al} (2012).}
\end{figure}

\begin{figure}[ht!]
  {\bf Embryonic stem cells (hESC) nuclei with random prepositioning of the chromosomes}\\
  \centering
  (a) \includegraphics[scale=0.4]{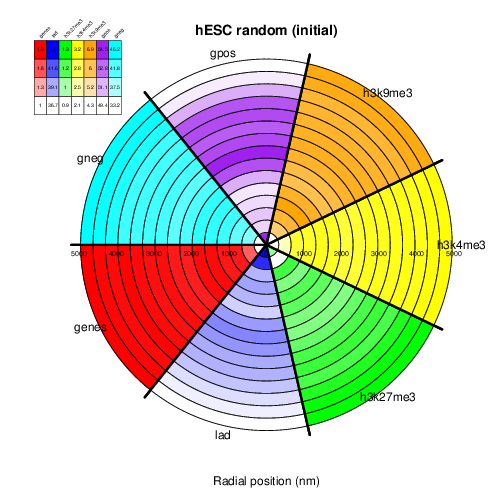}
  (b) \includegraphics[scale=0.4]{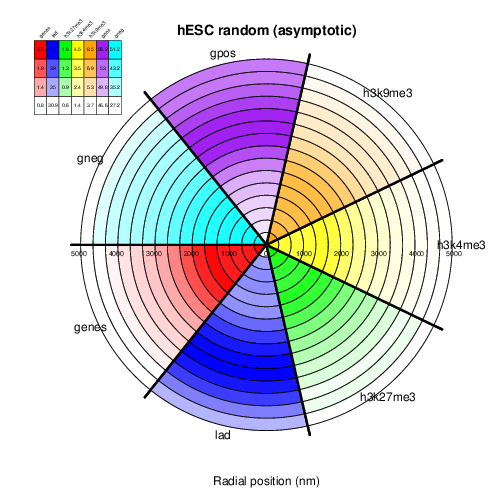}
    \caption{{\bf Nuclear positioning of functionally-oriented genomic regions.} Circular histogram giving the relative density (in percent) based on H3K9me3 (orange), H3K4me3 (yellow), H3K27me3 (green), LADs (blue) and genes (red), and negative (cyan) and positive (purple) Giemsa staining bands. Circle slices (thickness $\sim 320$nm) indicate radial position (in nm) within the bounding nucleus aggregated across all $10$ replicate simulations, and the numbers indicate the percentage of beads associated with the given feature relative to the total number of beads in the given circle slice. The figure is based on configurations obtained starting from a random prepositioning of the chromosomes, immediately prior to and after applying the steering protocol appropriate for hESC cell line based on the analysis of the data in Dixon \emph{et al} (2012).}
\end{figure}

\begin{figure}[ht!]
  \centering
  \includegraphics[width=\textwidth]{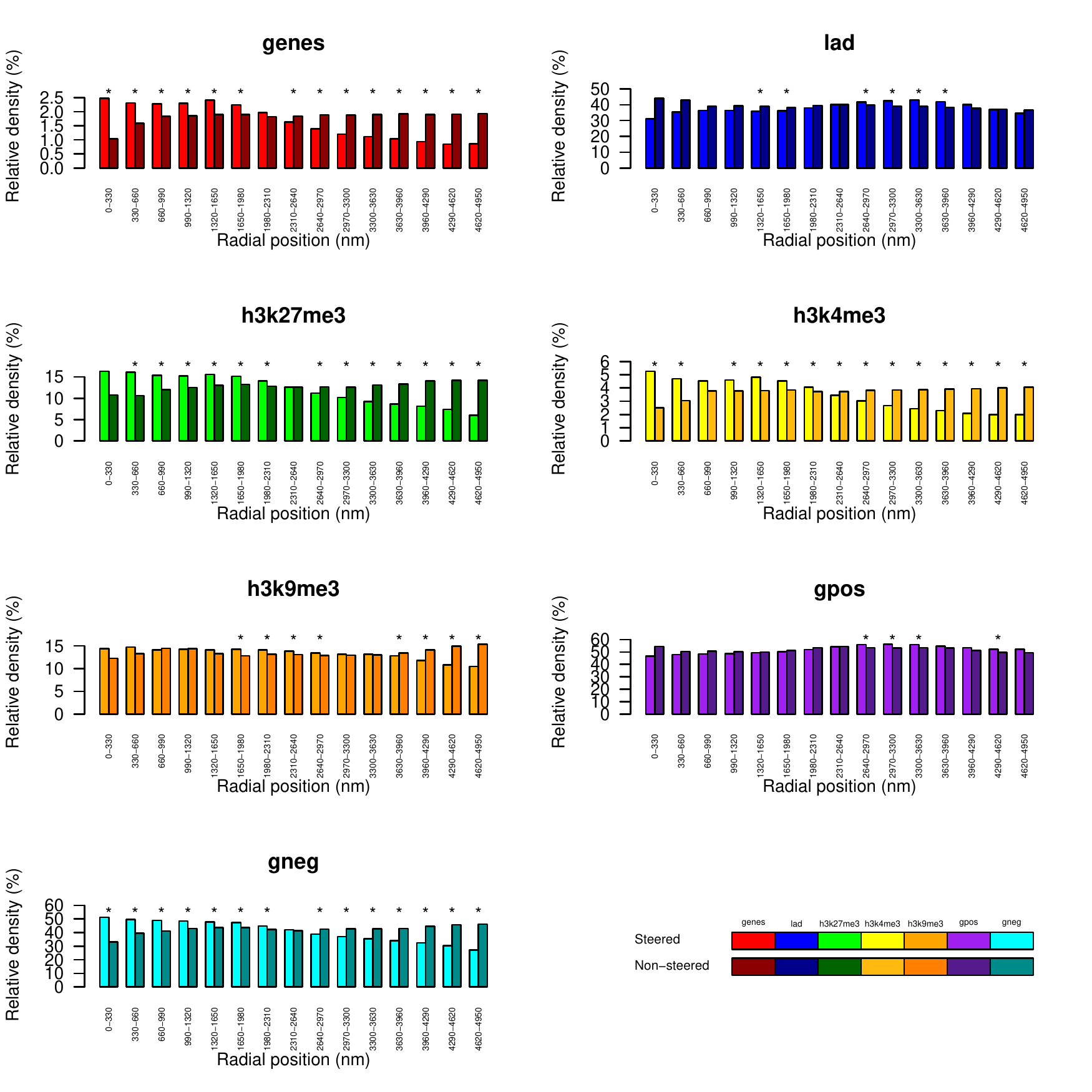}
  \caption{{\bf Histograms of the relative density of genes, LADs, H3K27me3, H3K4me3, H3K9me3, and positive (gpos) and negative (gneg) Giemsa staining bands in concentric equally thick radial shells of the nucleus.} The figure is based on configurations obtained starting from a random prepositioning of the chromosomes, immediately prior to (Non-steered) and after (Steered) applying the steering protocol appropriate for hESC cell line based on the analysis of the data in Dixon \emph{et al} (2012).}
\end{figure}


\clearpage

\begin{figure}[ht!]
  {\bf Lung fibroblasts (IMR90) nuclei with phenomenological prepositioning of the chromosomes}\\
  \centering
 (a) \includegraphics[width=0.4\textwidth]{./IMR90_initial_biased_std.png}
 (b) \includegraphics[width=0.4\textwidth]{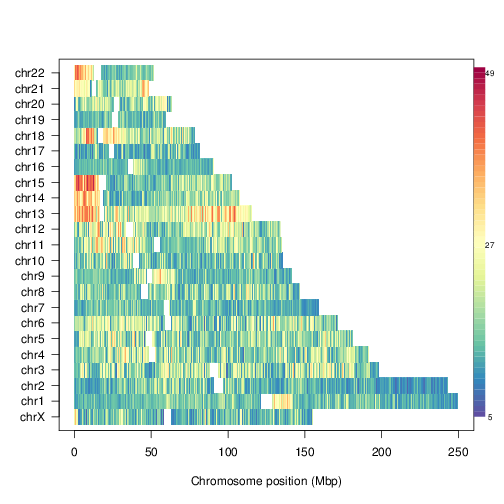}
  \caption{{\bf Genome-wide variability of radial bead position.} The panels are based on configurations obtained starting from the phenomenological prepositioning of the chromosomes in Bolzer {\em et al.} (2005), immediately prior to any steering (a) and after (b) applying the steering additional steering of the target pairs in Rao \emph{et al} (2014). Numbers indicate the standard deviation of the radial position across the $10$ replicate simulations.}
\end{figure}


\begin{figure}[ht!]
  {\bf Lung fibroblasts (IMR90) nuclei with phenomenological prepositioning of the chromosomes}\\
  \centering
(a)  \includegraphics[width=0.4\textwidth]{./IMR90_initial_biased_histogram.png}
(b)  \includegraphics[width=0.4\textwidth]{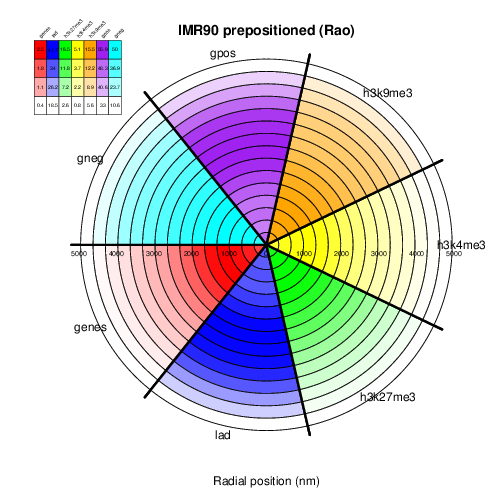}
  \caption{{\bf Nuclear positioning of functionally-oriented genomic regions.} Circular histograms giving the relative density (in percent) based on H3K9me3 (orange), H3K4me3 (yellow), H3K27me3 (green), LADs (blue) and genes (red), and negative (cyan) and positive (purple) Giemsa staining bands. Circle slices (thickness $\sim 320$nm) indicate radial position (in nm) within the nucleus aggregated across all $10$ replicate simulations, and the numbers indicate the percentage of beads associated with the given feature relative to the total number of beads in the given circle slice. The figure is based on configurations obtained starting from the phenomenological prepositioning of the chromosomes in Bolzer {\em et al.} (2005), immediately prior to any steering (a) and after (b) applying the additional steering of the target pairs in Rao \emph{et al} (2014).}
\end{figure}

\begin{figure}[ht!]
  \centering
  \includegraphics[width=\textwidth]{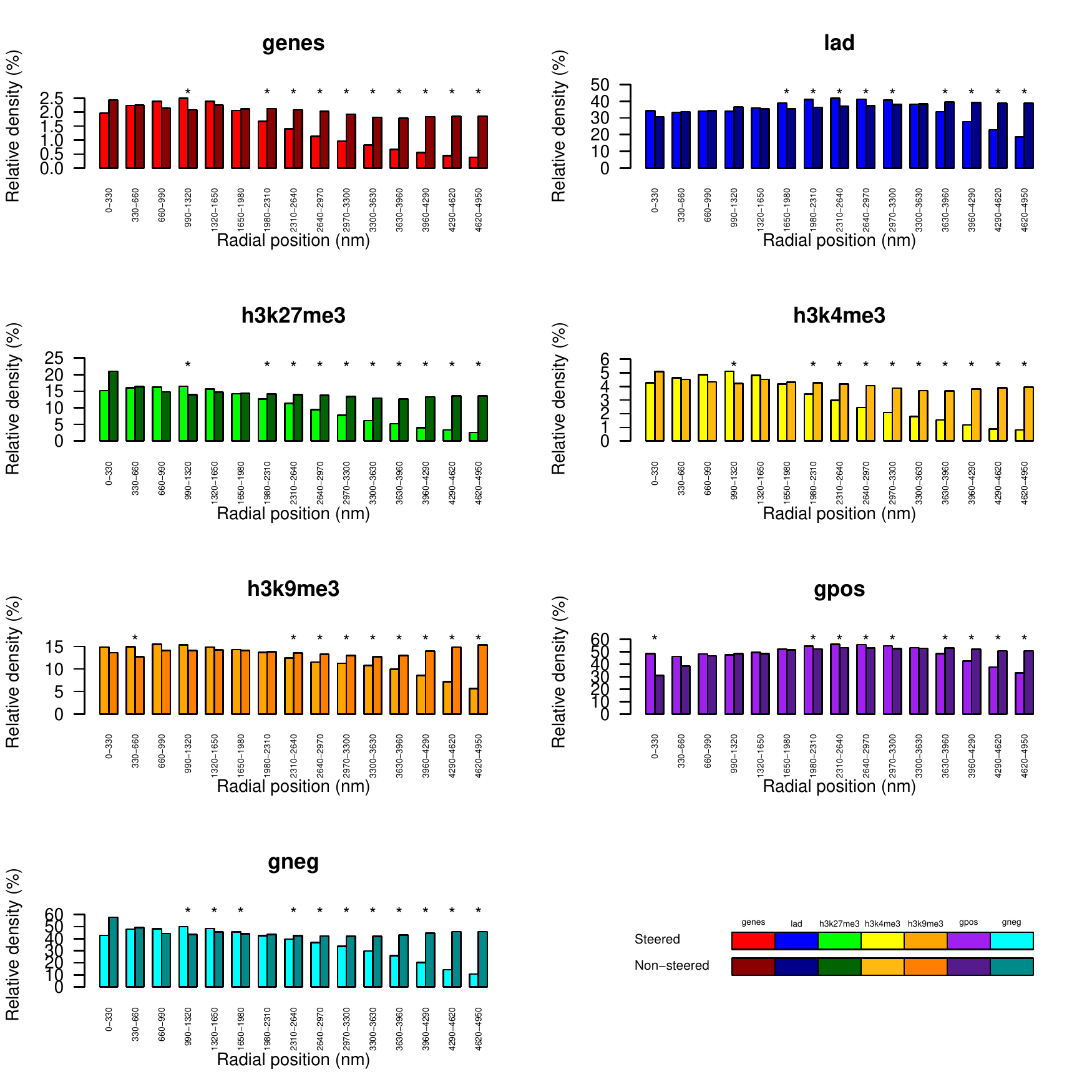}
  \caption{{\bf Histograms of the relative density of genes, LADs, H3K27me3, H3K4me3, H3K9me3, and positive (gpos) and negative (gneg) Giemsa staining bands in concentric equally thick radial shells of the nucleus.} The figure is based on configurations obtained starting from the phenomenological prepositioning of the chromosomes in Bolzer {\em et al.} (2005), immediately prior to any steering (Non-steered) and after applying the additional steering of the target pairs in Rao \emph{et al} (Steered). The asterisks indicate statistically significant differences between the two cases, using the Wilcoxon test with a p-value cutoff of $0.05$.}
\end{figure}


\begin{figure}[ht!]
  \centering
  \includegraphics[width=0.7\textwidth]{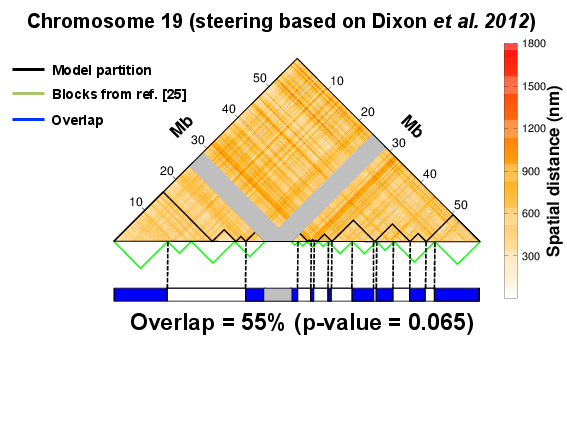}
  \caption{{\bf Structural macrodomain organization.} The upper triangle of the map of average spatial distance ($100$kb resolution) of chromosome 19 at the end of the steering dynamics based on the significant target contacts obtained from Dixon \emph{et al} (2012) is shown. The gray bands mark the centromeric region. The boundaries of the $13$ spatial macrodomains, identified with a clustering analysis of the distance matrix (see Methods), are overlaid on the map and the boundaries of the spatial \emph{blocks} in Kalhor {\em et al.} (2012) are shown below. The consistency of the two partitions is visually conveyed in the chromosome cartoon at the bottom. Overlapping regions, shown in blue, account for $55\%$ of the chromosome (centromere excluded).}
\end{figure}

\clearpage
\renewcommand{\figurename}{{\bf Supplementary Figure M}}
\setcounter{figure}{0}
\section*{Supplementary Methods}

\section{Modeling chromosome structure and dynamics} The feasibility to establish simultaneously the significant Hi-C contacts was explored by using model chromosomes and steered molecular dynamics simulations analogous to those used in ref.~\cite{distefano2013} and which are further detailed below.

\subsection{The chromosome polymer model} Each chromosome was modelled using a general bead-spring model~\cite{kremer_jcp}:
\begin{eqnarray}
  {\cal H} = U_{\rm LJ} + U_{\rm FENE} + U_{\rm KP} \text{.}
  \label{eq:Energy}
\end{eqnarray}
The first term is a truncated and shifted, purely repulsive Lennard-Jones potential:
\begin {equation}
  U_{LJ}(i,j)= \begin{cases}
    4k_{B}T \epsilon_{ij}\left[\left(\frac{\sigma}{d_{i,j}}\right)^{12}-\left(\frac{\sigma}{d_{i,j}}\right)^6+1/4\right]
    & \text{if $d_{i,j}\leq 2^{1/6}\sigma$},\\ 0 & \text{if
      $d_{i,j}>2^{1/6}\sigma$}{.}
  \end{cases}
  \label{eq:LJ}
\end{equation}
where $k_B$ is the Boltzmann constant, $T$ the temperature,
$\epsilon_{ij}$ is equal to $10$ if $\left|i-j\right|=1$, and $1$
otherwise, $\sigma=30$nm is the thickness of the chain and $d_{i,j}$
is the modulus of ${\vec d}_{i, j} = {\vec r}_i -
{\vec r}_j$ which is the distance vector between monomers $i$ and $j$ at positions ${\vec r}_i$ and
${\vec r}_j$, respectively. This term controls the \emph{cis}-chain excluded
volume interaction.

\noindent The second term is a FENE potential:
\begin {equation}
  U_{FENE}(i,i+1)=
  -150 k_B T \left( \frac{R_{0}}{\sigma} \right) ^{2}\left[1-\left(\frac{d_{i,i+1}}{R_0}\right)^2\right]
  \label{eq:FENE}
\end{equation}
where $R_0=1.5\sigma$ is the maximum bond length. This term ensures
the connectivity between consecutive beads of the same polymer
chain. The combined action of the FENE and LJ potential between consecutive beads is such that,
during the free and steered simulations, the average bonds length is close to $\sigma$ and never exceeds
$1.3\sigma$.

\noindent The third term is a Kratky-Porod, or bending potential:
\begin{equation}
  U_{br}(i,i+1,i+2)=\frac{k_{B}T\l_{p}}{\sigma}\left(1-\frac{\vec{d}_{i,i+1}\cdot\vec{d}_{i+1,i+2}}{d_{i,i+1}d_{i+1,i+2}}\right) ,
  \label{eq:KP}
\end{equation}
where the chain persistence length, $\l_{p}$, has been set equal to $5\sigma=150$nm to
reproduce the experimental rigidity of the chromatin fiber~\cite{bystricky}.

Different chromosomes interact only via excluded volume interactions, through the LJ repulsion of their constitutive beads.

\subsection{Description of the free chain dynamics}

The free dynamics of the chains was described with an underdamped Langevin equation, while the steering
process was guided by using pairwise harmonic constraints. In both cases the dynamics was integrated with the LAMMPS simulation package~\cite{lammps}.

Specifically, the underdamped Langevin equation is:
\begin{eqnarray}
  m \ddot{r}_{i \alpha} = - \partial_{i \alpha} {\cal H} -\gamma \dot{r}_{i \alpha} + \eta_{i \alpha} \left( t \right)
\end{eqnarray}
where is the $m$ is the bead mass which was set equal to the
LAMMPS default value, ${\cal H}$ is the system energy in
Eq.~\ref{eq:Energy}, the index $i$ runs over all the particles in the system,
and $\alpha = \left( x,y,z \right)$ indicates the Cartesian
components. The stochastic noise term $\eta_{i \alpha} \left( t \right)$ satisfies the
standard fluctuation dissipation conditions: $\langle \eta_{i
  \alpha} \rangle = 0$ and $\langle \eta_{i \alpha} (t) \, \eta_{j
  \beta} (t^\prime) \rangle = 2 \kappa_B {T} \gamma \, \delta_{ij}
\delta_{\alpha\beta} \delta(t-t^\prime)$, where
$\gamma=0.5\tau_{LJ}^{-1}$ is the friction coefficient, $\tau_{LJ} = \sigma (m/\epsilon)^{1/2}$ is the
Lennard-Jones time, $\delta_{ij}$ is the Kronecher delta, and
$\delta(t-t^\prime)$ is the Dirac delta.

The integration time step used in the LAMMPS numerical integration of the Langevin was equal to $\Delta t = 0.006 \tau_{LJ}$.

\subsection{Steered molecular dynamics protocol}
The colocalization of the target pairs of $100\,kbp$-long
chromosome stretches was promoted by using a steered molecular dynamics
protocol that progressively favoured the spatial proximity of the
target pairs in each model chromosome.

Specifically, we mapped each pair of selected regions, $A$ and $B$,
onto the corresponding $33$beads-long stretches of the chromosome
chain and added to the system energy an harmonic constrain:
\begin{eqnarray}
  U_{\rm H} = \frac{1}{2} k \left( L,t \right) (d_{A,B} - d_{0}/2.0)^ 2
\end{eqnarray}
\noindent where $d_{A,B}$ is the distance of the centers of mass of the
chromosome stretches and the equilibrium distance is set to half of
the contact distance $d_{0}=120$nm. The stiffness of the harmonic
constraint was controlled by a spring constant $k \left( L,t \right)$
depending on the sequence separation between the
two regions, $L$, and the simulation time, $t$.

The sequence-separation dependence of $k$ was introduced to ensure that the
steering process is not dominated by the target pairs at the largest sequence separation.
To this purpose, we made the spring
constant dependent on the sequence separation $L$ of the target pairs
so that, in the decondensed state (which is the state of the system
just before the steering process takes place) all pairs are pulled
together with the same average force, irrespective of their sequence
separation.

To accomplish this balancing of the spring constant, we used a
statistical reweighting approach.
Specifically, we computed the distribution of the square spatial
distances between all pairs of $100$kbp-long chromosome
stretches. We subdivided them in $25$ groups with the first, second,
etc. group gathering pairs at genomic distances, $L$, in the $0-10$Mbp,
$10-20$Mbp ranges, etc. Within each group, we
computed the normalized distribution of the spatial distances, $d$,
between al the pairs of $100$kbp chromosome strands in the
conformation of the decondensed chromosome system.
Each of the $6$ obtained distributions was fitted with the function:
\begin{eqnarray}
  y = A \cdot x^2 \cdot \exp \left( - \frac{x^2}{2 \, \sigma^2} \right)
\end{eqnarray}
which takes into account the radial weight. The distance distribution
and the Gaussian fit for bin $1$ is shown in
Supplementary Figure~\ref{fig:DistProbDistribution}A.

\begin{figure}[!ht]
\centering
A\includegraphics[width=0.4\textwidth]{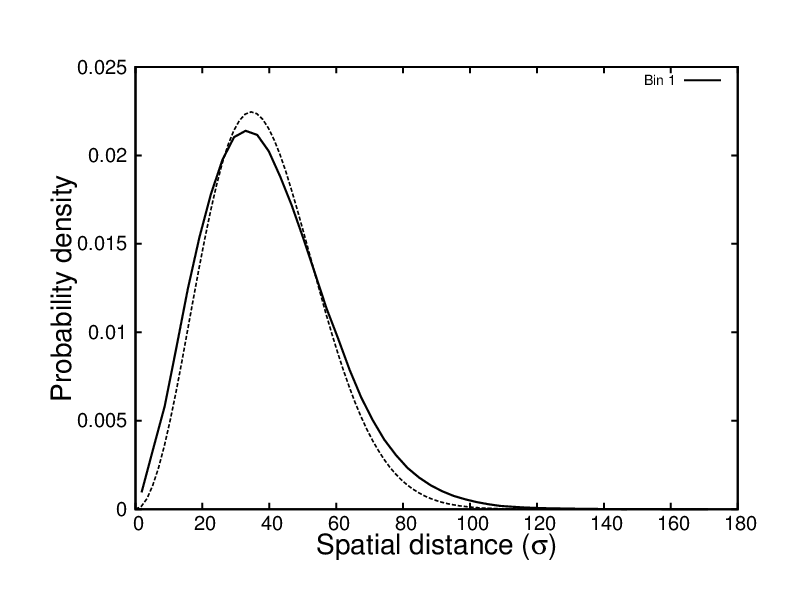}B\includegraphics[width=0.4\textwidth]{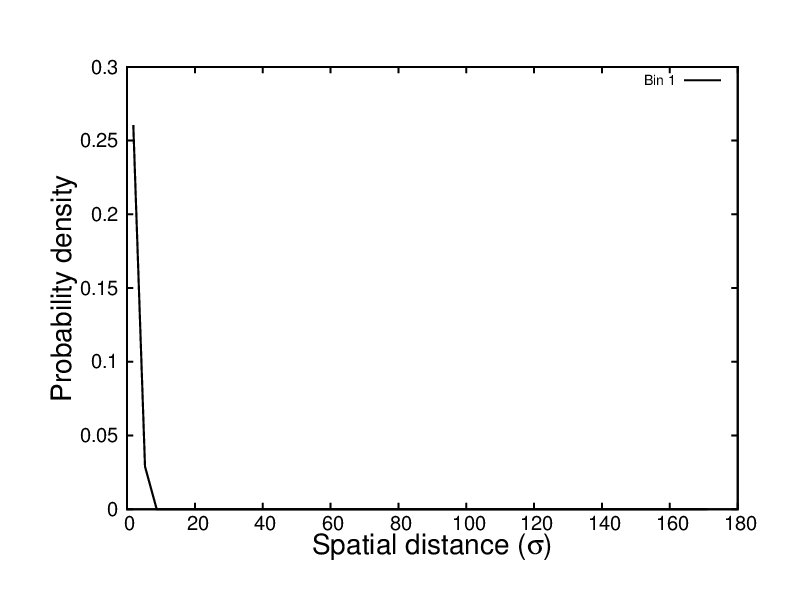}
\caption{Spatial distances probability density.}
\label{fig:DistProbDistribution}
\end{figure}

The gathered distance statistics was next reweighted with a Gaussian function:
\begin{eqnarray}
  w \left( k \right) = e^{-k/2(d - d_{0}/2)^2} \notag
\end{eqnarray}
where $d_0=120$nm is the cutoff distance for defining a contact
between target pairs. By doing so, we found the value of the spring
constant $k$ to yield at least $90.0\%$ of the pairwise distances
below the contact radius. The re-weighted probability distribution for
bin 1 is shown in Supplementary
Figure~\ref{fig:DistProbDistribution}B.

For robustness, the spring constant of pairs at separations larger than 60Mb (i.e. appreciably beyond the average chromosome length) was set equal to the same one of the pairs at distances in the 50-60Mb range
The obtained values of the reference spring constants are shown in Supplementary
Table~\ref{tab:spring_const} for each bin.

At the beginning of the simulation, the spring constants are set to $10\%$ of
these reference values and then are progressively increased.

\begin{table}
\centering
\begin{tabular}{|c|c|c|}
\hline
Bin & Genomic separation (Mb) & k \\
\hline
\hline
1    &  0-10  & 0.67714 \\
\hline
2    & 10-20  & 0.49177 \\
\hline
3    & 20-30  & 0.27520 \\
\hline
4    & 30-40  & 0.24356 \\
\hline
5    & 40-50  & 0.24399 \\
\hline
6    & 50-60  & 0.22315 \\
\hline
7-25 & 60-250 & 0.22315 \\
\hline
\end{tabular}
\caption{Spring constant.}
\label{tab:spring_const}
\end{table}

The simultaneous application of the $N_{tp}$ constraints to each
chromosome was implemented by using the PLUMED plugin for
LAMMPS~\cite{plumed}. The spring constants were gradually ramped
linearly every $6.0 \tau_{LJ}$ of steered simulation, so to avoid
driving the system significantly out of equilibrium: $k \left( L, t
\right) = k(L,0) t/(6.0 \tau_{LJ})$ for each value of $L$. Moreover,
the resulting pulling force between each constrained pair was
controlled every $6.0 \tau_{LJ}$ and, if it exceeded a maximum pulling
force of $300\,\epsilon/\sigma$, we set it to this maximum value. This
maximum corresponds to the nominal magnitude of the bonding force
(LJ+FENE) of the pairs of nearest neighbor beads at a distance of $1
\sigma$.

This maximum force was low enough and the simulation time-step $\Delta
t$ short enough to avoid appreciable over-stretching of the bond
connecting the beads, as this may result in unphysical passages of the
strands through each other during a numerical integration time step.

\subsection{The chromosome pre-mitotic re-condensation}
After steering, the model chromosomes were reconfigured to a linear,
pre-mitotic state. This was done by switching off the harmonic
restraints of the target contacts and by replacing them with restraints
between pairs of {\em loci} at the regular sequence separation of
$200$ kb. The new target constraints are applied using
harmonic springs having the same coupling constant equal to $2
\epsilon/\sigma^2$ during the entire simulation span. The equlibrium
distance of the springs is, instead, decreased in steps of $30$ nm
every $0.6$ $\tau_{LJ}$ from $200$ nm (the maximum extension of a
$200$ kb chromatin strand) to $30$ nm (the size of a bead). At the
last equilibrium distance, the simulation is prolungated up to $300$
$\tau_{LJ}$. This procedure results in the reconfiguration of the
chromosomes into a succession of $200$ kb loops arranged in a
string-like fashion. This target arrangement is analogous to the
mitotic or linear chromosome models of
refs.~\cite{sikorav1994,naumova2013}. This re-condesation procedure
is applied to the $10$ replicates of the human embryonic stem cells
(hESC) system before and after steering.

\subsection{Calculation of the contact distance for target pairs.}
\indent The bending properties of a polymer chain has a
large impact on the probability of looping, and hence, on bringing to
regions of the chain in spatial contact. The parameter that tunes this
property in the model used in this study is the persistence length
(see Eq.\ref{eq:KP}), which for a worm-like chain is half
the Kuhn-length, $l_K$, of the polymer chain.

\noindent To determine the contact radius for two constrained chromosome
stretches in a more general case (non-Gaussian chains), we considered
the expression of the mean square gyration radius $R_g^2$, which has
been established by Benoit and Doty in ref.~\cite{benoit1953} for a
worm-like chain of contour length $L_c$, which spans $M$ Kuhn-lengths
($M=L_c/l_K$):
\begin{eqnarray}
  \left< R^2_g \left( M \right) \right> = \frac{M \, l^2_k}{6} - \frac{l^2_k}{4} + \frac{l^2_k}{4 \, M} - \frac{l^2_k}{8 \, M^2} \left( 1 - e^{-2M} \right)
\label{eqn:RgBD}
\end{eqnarray}

A heuristic use of expression (\ref{eqn:RgBD}) is to estimate the
effective size of the region occupied in equilibrium by portions of
contour length $L_c$ from a long polymer chain. We therefore consider
the occupied region to be spherical, centred on the centre of mass of
the segment, and with a radius equal to $\sqrt{\left< R^2_g \left( M
    \right) \right>}$. The criterion to define an established spatial
contact between a pair of segments of contour length $L_c$ should be
based on the overlap volume of the two spheres spanned separately by the two stretches.

\noindent The volume of the intersection between two spheres of
identical radius $R$ as a function of the the distance $d$ between the
centres of the sphere is (see
\url{http://mathworld.wolfram.com/Sphere-SphereIntersection.html}):
\begin{eqnarray}
V = \frac{1}{12} \pi \left( 4R + d \right) \, \left( 2R - d \right)^2   \notag
\end{eqnarray}
As a criterion for a significant overlap, we consider the threshold of
$50\%$ of the volume of each individual sphere. The corresponding
sphere distance (contact radius) must be equal to:
\begin{eqnarray}
d & = & 0.694592710 \, R
  \notag
\end{eqnarray}

For the chromosome chains studied here, the contour length $L_c$ of
the stretches to co-localize accounts for $100$kbp which map onto $\sim
1 \mu m$~\cite{plospaper,bjpaper}. Given the Kuhn-length of the chromosome
fiber $l_K = 300\,nm$~\cite{plospaper,bjpaper}, $M$ results to be equal to
$\sim 3.3$. This corresponds to a contact distance of about:
\begin{eqnarray}
  d = 0.694592710 \sqrt{ \left< R^2_g \left( 3.3 \right) \right>} \sim 120 nm
\notag \end{eqnarray}

\end{document}